\documentclass[12pt,letterpaper]{article}
\pdfoutput=1
\usepackage{graphicx,array}
\usepackage{color}
\usepackage{latexsym}
\usepackage{amsthm}
\usepackage{amsmath}
\usepackage{amssymb}
\usepackage{hyperref}
\usepackage{bbold}

\def\simgt{\mathrel{\lower2.5pt\vbox{\lineskip=0pt\baselineskip=0pt
           \hbox{$>$}\hbox{$\sim$}}}}
\def\simlt{\mathrel{\lower2.5pt\vbox{\lineskip=0pt\baselineskip=0pt
           \hbox{$<$}\hbox{$\sim$}}}}

\newcommand{\be}{\begin{equation}}
\newcommand{\ee}{\end{equation}}
\newcommand{\bea}{\begin{eqnarray}}
\newcommand{\eea}{\end{eqnarray}}
\newcommand{\Eq}[1]{Eq.~(\ref{#1})}

\newcommand{\Sec}[1]{Sec.~\ref{#1}}

\newcommand{\Fig}[1]{Fig.~(\ref{#1})}

\newcommand{\App}[1]{App.~\ref{#1}}

\newcommand{\OO}{\mathcal{O}}

\hypersetup{colorlinks,citecolor= blue,linkcolor= black}

\begin{document}

\hfill

\vspace{1cm}

\begin{center}
{\LARGE\bf
Prospects and
Blind Spots for \\
\vspace*{0.5cm}
Neutralino Dark Matter
}\\
\bigskip\vspace{1cm}{
{\large Clifford Cheung$^1$, Lawrence J. Hall$^2$, 
David Pinner$^2$, and Joshua T. Ruderman$^2$}
} \\[7mm]
 {\it 
$^1$California Institute of Technology, Pasadena, CA 91125 \\
 $^2$Berkeley Center for Theoretical Physics, University of California, and\\
   Theoretical Physics Group, Lawrence Berkeley National Lab,  Berkeley, CA 94720} \end{center}
\bigskip
\centerline{\large\bf Abstract}

\vspace{0.25cm}

Using a simplified model framework, we assess observational limits and discovery prospects for neutralino dark matter, taken here to be a general admixture of bino, wino, and Higgsino. Experimental constraints can be weakened or even nullified in regions of parameter space near 1) purity limits, where the dark matter is mostly bino, wino, or Higgsino, or 2) blind spots, where the relevant couplings of dark matter to the $Z$ or Higgs bosons vanish identically. We analytically identify all blind spots relevant to spin-independent and spin-dependent scattering and show that they arise for diverse choices of relative signs among $M_1$, $M_2$, and $\mu$. At present, XENON100 and IceCube still permit large swaths of viable parameter space, including the well-tempered neutralino. On the other hand, upcoming experiments should have sufficient reach to discover dark matter in much of the remaining parameter space. Our results are broadly applicable, and account for a variety of thermal and non-thermal cosmological histories, including scenarios in which neutralinos are just a component of the observed dark matter today. Because this analysis is indifferent to the fine-tuning of electroweak symmetry breaking, our findings also hold for many models of neutralino dark matter in the MSSM, NMSSM, and Split Supersymmetry. We have identified parameter regions at low $\tan \beta$ which sit in a double blind spot for both spin-independent and spin-dependent scattering. Interestingly, these low $\tan \beta$ regions are independently favored in the NMSSM and models of Split Supersymmetry which accommodate a Higgs mass near 125 GeV.

\begin{quote} \small

\end{quote}

\newpage

\tableofcontents
\newpage
\section{Introduction}
\label{sec:intro}

In many supersymmetric theories, the lightest superpartner (LSP) is a stable neutralino which can account for the observed dark matter (DM) in the Universe.  However,  two complementary experimental efforts seemingly cast doubt on this possibility, at least 
for the simplest case where the LSP is a linear combination of the bino, wino, and Higgsino, $\chi \sim (\tilde{b}, \tilde{w}, \tilde{h}$).

First, a variety of searches from the Large Hadron Collider (LHC) have placed compelling limits on supersymmetry, with constraints especially stringent when the LSP is a stable, weakly interacting massive particle (WIMP).  Naively, this casts doubt on the so-called ``WIMP miracle.''
However, the powerful null results from the LHC apply to colored superpartners decaying to missing energy---direct limits on the bino, wino, and Higgsino still remain weak.
So while the LHC challenges supersymmetry as a solution to the hierarchy problem, it does not impose strong, direct, constraints on the origin of DM.

Secondly, after decades of improving technologies and increasing target masses, the direct detection of galactic DM has reached unprecedented levels of sensitivity, as shown in \Fig{fig:DirectDetectLim}.  At present, the best limits on the scattering of DM against target nuclei are from the XENON100 experiment \cite{Aprile:2012nq}, which probes spin-independent (SI) and spin-dependent (SD) scattering.  Complementary and in some cases more powerful constraints on SD scattering have also been obtained by the IceCube observatory~\cite{IceCubeTalk}, which searches for high energy neutrino signals originating from DM accumulating inside the sun.  
In many theories---for example neutralino DM---SI and SD scattering is mediated by Higgs and $Z$ boson exchange, respectively.  Cross-sections corresponding to different values of the couplings $c_{ h\chi\chi}$ and $c_{Z \chi\chi}$ are shown in \Fig{fig:DirectDetectLim}, which reflect the fact that $\sigma_{\rm SI} \propto c_{h \chi \chi}^2$ and $\sigma_{\rm SD} \propto c_{Z \chi \chi}^2$.  For neutralino DM, both couplings originate from the electroweak gauge couplings $g' \sim 0.35$ and $g \sim 0.65$, so the naive conclusion of \Fig{fig:DirectDetectLim} is that neutralino DM is presently excluded.  However, this argument against neutralino DM is incorrect.  The couplings $c_{h \chi\chi}$ and $c_{Z \chi\chi}$ arise from multiple contributions of the same order which can constructively or destructively interfere; upon squaring the resultant couplings, one finds SI and SD cross-sections which can easily be enhanced or suppressed by an order of magnitude from the naive expectation.  While current constraints do not require any particular fine-tuning among parameters---just relative signs---this will not be so easy for future limits.

In this paper, we explore the observational status and future of neutralino DM in the context of simplified models, defined here to be
minimal theories of weak scale SUSY, decoupling all but a handful of superpartners relevant for DM phenomenology.  Crucially, these models are characterized by a small number of theory parameters defined at the weak scale.
Our aim is to identify the regions of parameter space that are presently allowed and understand the detection prospects for upcoming experiments.

A priori, the parameter space of supersymmetric DM is vast, which is why typical analyses employ exhaustive parameter scans.  However, the majority of these parameters are irrelevant to neutralino DM if we assume that the scalar states---heavy Higgs bosons, squarks, and sleptons---are sufficiently decoupled to not significantly affect processes relevant to the cosmological or observational properties of DM\@.   This simplified model approach is further motivated by the absence so far of signals at the LHC for supersymmetry and heavy Higgs states.  Similarly, LHC data motivate the assumption that the SI scattering of DM against the nucleon is dominated at tree-level by the exchange of a standard model (SM) Higgs of mass near 125 GeV\@.
These assumptions lead to a manageable parameter space for neutralino DM which an admixture of bino, wino, and Higgsino.  This parameter space is comprised of mass parameters for $\tilde{b}, \tilde{w}$ and  $\tilde{h}$ and the ratio of vacuum expectation values,
\bea
\chi \sim (\tilde{b}, \tilde{w}, \tilde{h})\quad &:& \quad  (M_1, M_2, \mu, \tan \beta).
\eea  
We find that our analysis is little altered for squarks as light as 1 TeV, and we quantify their effects for lower mass values.  For simplicity, throughout this work we assume CP conservation and we consider $m_\chi \gtrsim m_W$, so that DM is above threshold to annihilate into $W^+W^-$.  In particular, we do not consider the $Z/h$ pole regions, and we do not consider light DM, $m_\chi \lesssim 10$~GeV.

This minimal framework for neutralino DM is relevant for a remarkably wide range of theories.  The most obvious case is the minimal supersymmetric standard model (MSSM) with large $\tan \beta$ and multi-TeV squarks to yield a 125 GeV Higgs boson.  On the other hand, $\chi$ can also describe neutralino DM in the next-to-minimal supersymmetric standard model (NMSSM), provided that the singlino component of $\chi$ is sufficiently small.  Such theories can account for a 125 GeV Higgs boson with less fine-tuning in electroweak symmetry breaking than the MSSM, and prefer smaller values of $\tan \beta$~\cite{Hall:2011aa}.  
Our setup also characterizes DM in a variety of theories of Split Supersymmetry \cite{ArkaniHamed:2004fb}, with $\tan \beta$ typically decreasing as the mass of the scalar superpartners is raised.  In Split Supersymmetry, the decoupling of the scalars is guaranteed, while in other theories this is a greater assumption.  
 Notably, the existence of light DM does not invoke any additional tuning, since the neutralino mass is protected by chiral symmetries.    In this paper we are agnostic about the role of  supersymmetry for the naturalness puzzle---as such, our analysis applies to all of the above theories.  

Our main conclusions are as follows:
\begin{itemize}
\item Despite stringent SI and SD limits, neutralino DM remains experimentally viable.  The allowed regions of parameter space are large; they permit thermal, non-thermal, and multi-component neutralino DM, and include points with minimal fine-tuning.  This result hinges on the proximity of viable parameter regions to numerous direct detection ``blind spots'' at which the SI or SD scattering cross-section vanish identically due to destructive interference in the couplings $c_{h\chi\chi}$ or $c_{Z \chi\chi}$.  We have analytically identified all such cancellation points for neutralino DM and found that they require diverse sign choices among the mass parameters (summarized in Table~\ref{tab:blindspots}).   
\item
Upcoming direct detection experiments will probe the bulk of neutralino DM parameter space, except for regions very close to the direct detection blind spots.  Most of the regions left unscathed require fine-tuning between $M_1$, $M_2$, and $\mu$ in order to track the blind spots.  A notable exception to this is thermal bino/Higgsino and non-thermal Higgsino DM at low $\tan \beta$, which will evade both SI and SD experiments for the foreseeable future.  Interestingly, this region of parameter space is theoretically favored: low $\tan \beta$ is required by natural theories like $\lambda$SUSY~\cite{Barbieri:2006bg} in order to sufficiently boost the Higgs mass; it is also required by unnatural theories like Split Supersymmetry \cite{ArkaniHamed:2004fb}, given Higgs mass constraints.
\item
Bino/wino DM remains an attractive candidate for neutralino DM\@.  As discussed in past works \cite{BirkedalHansen:2001is,BirkedalHansen:2002am,Baer:2005zc,Baer:2005jq,ArkaniHamed:2006mb}, the observed DM abundance can be accommodated by thermal freeze-out through coannihilations.  We present here the first systematic analysis of the full parameter space of thermal, non-thermal, and multi-component bino/wino DM in relation to present and future experiments.
\end{itemize}

Let us comment briefly on the relationship between our results and past work.  Two recent studies of neutralino DM~\cite{Grothaus:2012js, Perelstein:2012qg} focus on the correlation between the size of the cross-section relevant for direct detection and naturalness in electroweak symmetry breaking.  
An earlier study~\cite{Mandic:2000jz} identified certain regions of parameter space with small direct detection cross-section.  We note that these earlier papers all relied on scans over large parameter spaces,  differing greatly from our approach of using simplified models.
Well-tempered bino-Higgsino DM was studied in a simplified model approach in \cite{Farina:2011bh}, and in the addendum it is claimed that the recent XENON100 results exclude the case of thermal freeze-out.
However, the case of a relative sign between $\mu$ and $M_1$, leading to destructive interference in the Higgs coupling $c_{h \chi \chi}$, was not considered by these authors.  

The outline of our paper is as follows. In \Sec{sec:obs}, we discuss the relevant experiments, focusing on present limits and future reach to probe the SI and SD cross-sections.  We go on to discuss the cosmological history of the DM relic abundance in \Sec{sec:relic}, including a review of the well-tempered neutralino.  In \Sec{sec:scattsupp}, we identify regions of parameter space where the DM direct detection cross-section is suppressed.   This suppression can come from purity of DM, which we discuss in \Sec{subsec:purity}, or from blind spots, where the Higgs or $Z$ coupling vanish, which we classify in \Sec{subsec:blindspots}.  We present detailed results on the present limits and reach for the simplified model where DM is a mixture of a bino and a Higgsino, in \Sec{sec:bh}.  Then we bring the wino into the spectrum in \Sec{sec:bw}, and study DM that is a mixture of bino, wino, and Higgsino (with a special focus on the case where DM is dominantly bino/wino).  Our conclusions appear in \Sec{sec:concl}.

We have included a few appendices with results that are of a more technical nature, but important for our study.  Direct detection limits depend on the strange quark content of the nucleon, $f_s$, and we use the lattice values~\cite{Giedt:2009mr} for our analysis.   In App.~\ref{app:fs}, we review the recent status of determinations of $f_s$ and quantify how our results are sensitive to this quantity.  In App.~\ref{app:tune}, we explain how we quantify the tuning of the DM abundance and direct detection cross-section, independently of a possible tuning of the electroweak scale.  Throughout the paper, we use tree-level scattering cross-sections, and one may wonder how our limits, and specifically blind spots, are modified by loop corrections.  In App.~\ref{app:loop} we justify that these loop corrections should be small throughout most of the parameter space we consider.

\section{Observational Constraints}
\label{sec:obs}

In this section, we enumerate the experiments relevant to neutralino DM and broadly outline their status and future reach.
However, before delving into the experimental limits we would like to ask: what is the characteristic size for the SI and SD cross-sections expected of neutralino DM which couples through the Higgs and Z bosons?
Given the interactions, 
\be
{\cal L} \supset \frac{c_{h \chi \chi}}{2} \,\, h ( \chi \chi  +  \chi^\dagger \chi^\dagger) 
\;+\; c_{Z \chi \chi} \, \chi^\dagger  \bar \sigma^{\mu} \chi Z_{\mu},
\ee
then in the limit in which the DM is heavier than the nucleon, the SI and SD cross-sections are
\be
\sigma_{\rm SI} = 8 \times 10^{-45}~\mathrm{cm}^2 \left(  \frac{c_{h \chi \chi}}{0.1}\right)^2 
\hspace{1in}
\sigma_{\rm SD} = 3 \times 10^{-39}~\mathrm{cm}^2 \left( \frac{c_{Z \chi \chi} }{0.1}\right)^2.
\ee
While $\sigma_{\rm SD}$ is typically considerably larger than $\sigma_{\rm SI}$, SI experimental constraints are commensurately stronger than SD, so these two limits are comparable in strength~\cite{Belanger:2008gy,Cohen:2010gj}.
Note that $\sigma_{\rm SI}$ depends on nuclear form factors, in particular the strange quark content of the nucleon.  For our analysis we adopt the lattice values of \cite{Giedt:2009mr}.  A more technical discussion of the strange quark content of the nucleon is contained in App.~\ref{app:fs}. 

\begin{figure}[t]
\begin{center} 
\includegraphics[scale=.45]{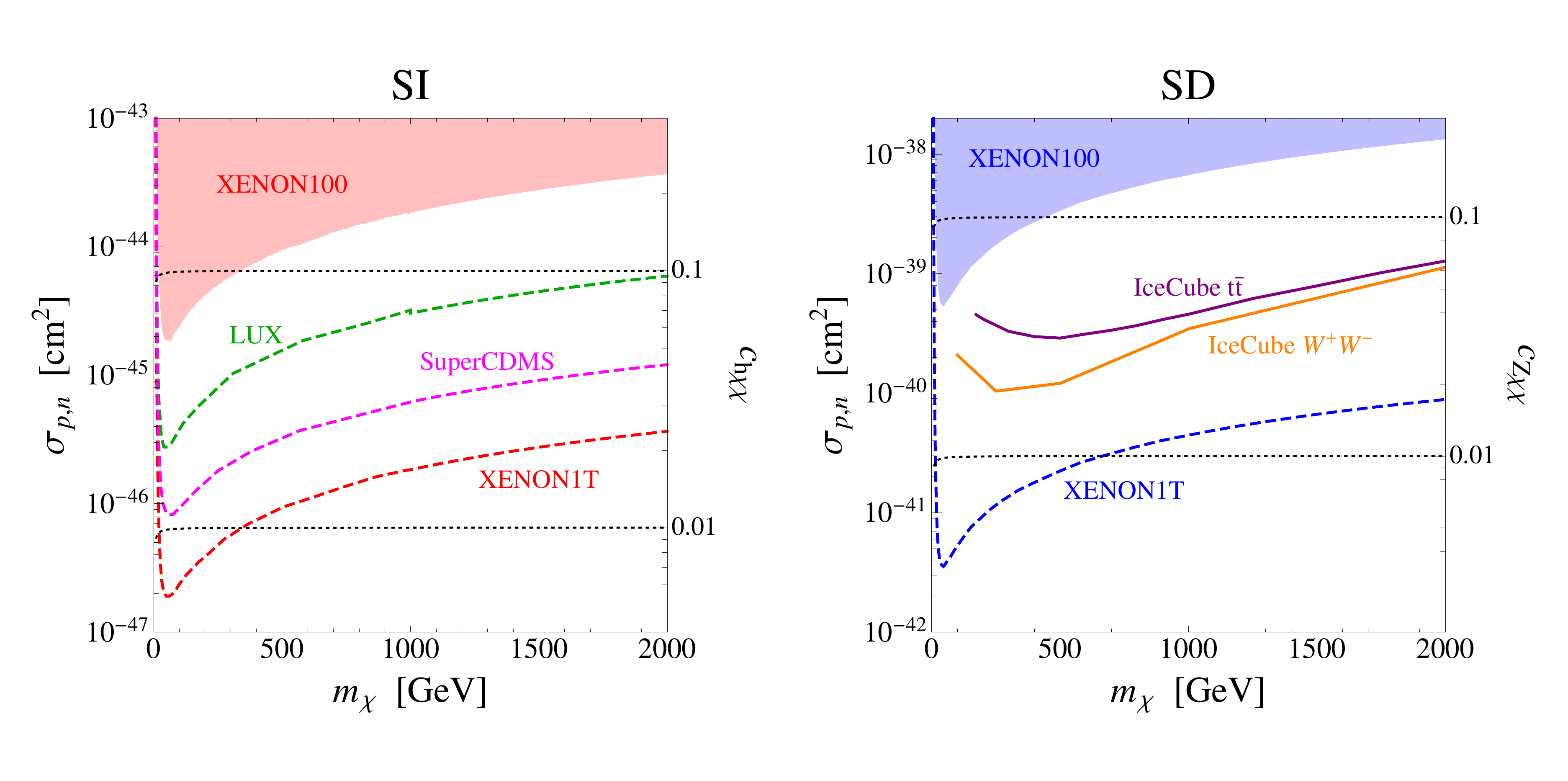}
\end{center}
\caption{ \label{fig:DirectDetectLim}
Present limits (filled or solid) and future reach (dashed) for SI/SD scattering of DM, shown in terms of the cross-section (left axis) or DM Higgs/$Z$ coupling (right axis).  For SI scattering we show the current limit from XENON100~\cite{Aprile:2012nq} as well as the projections for LUX~\cite{Akerib:2012ak}, SuperCDMS~\cite{SuperCDMSTalk}, and XENON1T~\cite{Aprile:2012zx}.  
For SD scattering we show the current limit from XENON100~\cite{XenonSDtalk} on DM-neutron scattering, as well as the current limit from IceCube~\cite{IceCubeTalk} on DM-proton scattering, assuming annihilations into $W^+ W^-$ or $t \bar t$ (estimated).  We also show our estimate for the reach of XENON1T~\cite{LangChat} for DM-neutron scattering.  
}
\end{figure}

The SI scattering of DM with nucleons is highly constrained by null results from direct detection experiments.  At the forefront of this experimental effort is XENON100~\cite{Aprile:2012nq}, an underground, two-phase DM detection experiment which employs a 62 kg radio-pure liquid Xe target.  As shown to the left of \Fig{fig:DirectDetectLim}, XENON100 provides the current leading experimental limit on SI scattering; their latest limit uses an exposure of 0.02 ton~$\times$~years.  Throughout this paper, we present 90\% C.L. limits and reach, and we take the local DM density to be $\rho_0 = 0.3~\mathrm{GeV}/\mathrm{cm}^2$, which is supported by a recent direct measurement using stellar kinematics, $\rho_0 = 0.3 \pm 0.1~\mathrm{GeV}/\mathrm{cm}^2$~\cite{Bovy:2012tw}.  We do not attempt to incorporate astrophysical uncertainties into our analysis. 

XENON100  will be far surpassed by XENON1T~\cite{Aprile:2012zx}, which is projected to begin collecting data in $\sim 2015$ and  should offer substantially better sensitivity due to a larger  Xe target mass of  $\sim$ 2.2 tons.  The projection in \Fig{fig:DirectDetectLim} shows the estimated limit with an exposure of 2.2 ton~$\times$~years.  Meanwhile, LUX~\cite{Akerib:2012ak}---a direct detection experiment of similar design but with a Xe target volume of 350 kg---is slated for operation in $\sim 2013$, and has a projected reach between that of XENON100 and XENON1T\@.   In \Fig{fig:DirectDetectLim} we show the conservative LUX reach estimate from Ref.~\cite{Akerib:2012ak}, which assumes an exposure of 0.08 ton~$\times$~years and a light collection efficiency of 15\%.  A more optimistic light collection efficiency of 20\% improves the limit by a factor of $\sim1.6$.
We also show the projected limit from SuperCDMS at SNOLAB~\cite{SuperCDMSTalk}, which is based on a complementary technology utilizing germanium detectors.  Throughout the rest of the paper, we will focus on the XENON100 limit, the conservative LUX estimate, and the XENON1T reach, but it is understood that the SuperCDMS reach would fall between the LUX and XENON1T estimates.  We note that ton-scale Xenon detectors are also being pursued by LZ~\cite{Malling:2011va} and PandaX~\cite{PandaXTalk} and ton-scale Argon detectors are being pursued by the DarkSide collaboration~\cite{Wright:2011pa}; we do not show their reach but our XENON1T curves should be taken as representative of the expected sensitivity of ton-scale liquid noble gas detectors.

The SD scattering of DM with nucleons is constrained by direct detection experiments.  The right of \Fig{fig:DirectDetectLim} shows the preliminary limit from XENON100, with 100 livedays, on the DM-neutron scattering cross-section~\cite{XenonSDtalk}.    
We assume a nuclear shell model that leads to a conservative limit; a different model improves the limit by a factor of $\sim 1.9$, and this difference can be taken as an estimate of the uncertainty on the limit from nuclear physics.
We also show an estimate of the reach with XENON1T\@.  No official SD reach estimate has been released by the XENON1T collaboration, so we estimate the reach by rescaling the XENON100 limit  by the expected difference of exposure between the XENON100 limit (100.9 days~$\times$~48 kg) and XENON1T (2.2 ton~$\times$~years)~\cite{LangChat}.

There is also a constraint on SD DM-proton scattering from neutrino telescopes, which probe the annihilation into neutrinos of DM captured in the sun. We show the preliminary limit from IceCube~\cite{IceCubeTalk} in \Fig{fig:DirectDetectLim}, utilizing 79 strings and 317 days livetime.  The limit from IceCube is a function of the neutrino spectrum, which depends on the DM annihilation products.  We show the limit from DM annihilations into $W^+ W^-$, which is released by the IceCube collaboration, as well as our estimate of the limit  if DM annihilates entirely into $t \bar t$.  In order to perform this estimate, we use the IceCube $W^- W^+$ limit at fixed DM mass to determine the muon flux limit for mono-energetic $W$'s~\cite{Wikstrom:2009kw}.  We determine the $W$ energy spectrum resulting from top decays from annihilations at each DM mass using {\tt{MadGraph}}~\cite{Alwall:2011uj}, and estimate the upper limit by approximating the upper limit on the number of observed muons to be independent of the $W$ energy.

We briefly comment on a few relevant constraints on neutralino DM other than direct detection.  There are indirect limits on DM annihilations into gamma rays.  The strongest constraint comes from a combined Fermi analysis of 10 satellite galaxies using 2 years of data~\cite{Ackermann:2011wa}.  DM annihilating into $W^- W^+$ is constrained to have a cross-section smaller than $\left< \sigma v \right> \lesssim 10^{-25}~\mathrm{cm}^3/\mathrm{s}$, which as we will see places important constraints on DM with a non-thermal cosmology. The limit includes uncertainties on the haloes of the satellite galaxies, and should be viewed as conservative with regard to these uncertainties.  In principle anti-proton measurements from PAMELA may set complementary limits~\cite{Adriani:2010rc}; however, we restrict our focus to Fermi because gamma rays, unlike antiprotons, are not sensitive to propagation uncertainties.  There are also a few relevant limits coming from colliders.  LEP2 constrains the charged components of Higgsinos and winos: $m_{\chi^\pm} \gtrsim 100$~GeV~\cite{LEP}.  There are now limits from the LHC constraining winos lighter than $\sim300$~GeV that decay to a neutralino lighter than about 100~GeV~\cite{ATLASinos, CMSinos}.  We will not consider the LHC limits further in this paper because we focus on DM heavier than 100 GeV, where these limits are not relevant.

\section{Relic Abundances and Well-Tempering}
\label{sec:relic}

WMAP observations are consistent with a relic abundance of DM given by~\cite{Beringer:1900zz}:
\bea
\Omega_{\rm obs} h^2 = 0.111 \pm 0.006\;  (1 \sigma).
\label{eq:omegaobs}
\eea
Throughout our analysis,  $\Omega_\chi$ denotes the total relic abundance of neutralino DM, while $\Omega_{\chi}^{(\textrm{th})}$ denotes the relic abundance of neutralino DM expected from thermal freeze-out alone.   To be comprehensive, our analysis accommodates three scenarios for the cosmological history: 
\begin{itemize}
\item
{\bf Thermal} ($\Omega_{\rm obs}=\Omega_{\chi} = \Omega_{\chi}^{(\rm th)}$).  DM is solely comprised of neutralinos arising from thermal freeze-out.
\item {\bf Non-Thermal} ($\Omega_{\rm obs} =\Omega_{\chi} \neq \Omega_{\chi}^{(\rm th)}$).  DM is solely comprised of neutralinos, but thermal freeze-out either over- or under-produces.  We assume that non-thermal processes either deplete or enhance the abundance to exactly saturate the WMAP constraint.
\item {\bf Multi-Component} ($\Omega_{\rm obs} > \Omega_{\chi} = \Omega_{\chi}^{(\rm th)}$).  DM is partly comprised of neutralinos arising from thermal freeze-out.  We assume that the balance of DM is provided by a secondary DM particle, {\it e.g.}~axions.
\end{itemize}

In the second and third cases, $\Omega_\chi^{(\rm th)}$ is not stringently constrained by WMAP measurements, so these scenarios offer greater freedom for evading experimental constraints.  In the first case, however, the relic abundance is fixed to the observed WMAP value, and for $\chi$ DM this typically requires a modest fine-tuning among parameters.  This occurs because pure bino DM is over-abundant, while pure wino or Higgsino DM is under-abundant for masses below 1 TeV \cite{Cirelli:2005uq,Cirelli:2007xd} and 2.7 TeV \cite{Cirelli:2007xd,Hisano:2006nn}, respectively. Thus, only a precise admixture of bino and wino or Higgsino---{\it i.e.}~a well-tempered neutralino---can accommodate $\Omega_{\rm obs} = \Omega_{\chi}^{(\rm th)}$ \cite{ArkaniHamed:2006mb} (for earlier refs., see~\cite{Feng:2000gh,Giudice:2004tc,Pierce:2004mk}).     

\begin{figure}[t]
\begin{center} 
\includegraphics[scale=.49]{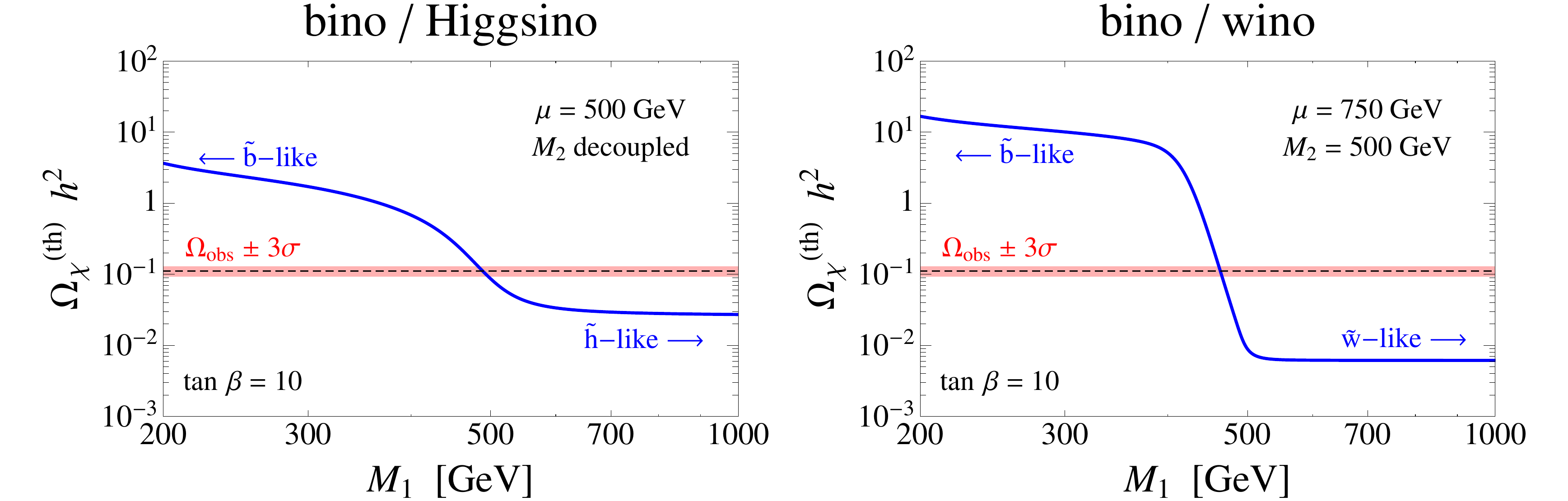}
\end{center}
\caption{\label{fig:TemperOmega}
The thermal freeze-out abundance for $\chi \sim (\tilde{b}, \tilde{h})$ (left) and $\chi \sim (\tilde{b}, \tilde{w})$ (right) is especially sensitive to  parameters near the well-tempered cross-over region. The relic abundance is exponentially sensitive in bino/wino DM, where thermal freeze-out follows mostly from coannihilation. 
}
\end{figure}
\Fig{fig:TemperOmega} shows the dependence of $ \Omega_{\chi}^{(\rm th)}$ on $M_1$ for the cases of $\chi \sim (\tilde{b}, \tilde{h})$ and $\chi \sim (\tilde{b}, \tilde{w})$.  Here and throughout the paper, we compute relic densities with {\tt MicrOMEGAs~2.4.5}~\cite{Belanger:2006is}.
We do not  include the effect of Sommerfeld enhancement, a non-perturbative effect which can substantially boost the annihilation cross-section of DM if it is much heavier than a force carrying particle.  Sommerfeld enhancement through electroweak bosons is an especially important effect for wino DM $\gtrsim 2$ TeV, which is not our focus.  The parameters in \Fig{fig:TemperOmega} have been chosen so that in the limit of heavy $M_1$,  $\chi$ is dominantly a Higgsino or wino with mass 500 GeV\@.   At low $M_1$, $\chi$ is dominantly bino, and as $M_1$ increases it gradually acquires a larger component of Higgsino/wino allowing it to annihilate more rapidly to final states involving $W$, $Z$, and $h$.  As $M_1$ approaches 500 GeV the abundance changes rapidly---partly because the mixing angle changes rapidly and partly because the LSP mass is approaching the mass of the next lightest neutralino and chargino states, allowing for coannihilation \cite{Griest:1990kh}.  As is well-known, coannihilation is exponentially sensitive the mass difference between the DM and its neighboring states.  The effective freeze-out cross-section for $i = 1, 2, \ldots, N$ states is given by
\bea
\langle \sigma v \rangle_{\rm coann} &=&  \frac{\sum_{i,j}^N w_i w_j \sigma_{ij} x^{-n}}{\left(\sum_i w_i \right)^2} \\
w_i &=&  \left(\frac{m_{\chi_i}}{m_\chi} \right)^{3/2}e^{-x(m_{\chi_i}/m_\chi -1)},
\label{eq:coann}
\eea
where $x= m_\chi/T$ and $m_{\chi_1}\equiv m_\chi$.  Coannihilation  dominates the transition region in the bino/wino case, leading to a curve that is much steeper than for the bino/Higgsino case.  This is the consequence of the exponential sensitivity to mass parameters in \Eq{eq:coann}.

Because thermal neutralino DM requires a special relation among parameters, $|\mu| \approx |M_1|$ or $|M_2|$, as we see in \Fig{fig:TemperOmega}, it is worthwhile to pause and consider our philosophy on fine-tuning.  In this paper, we are agnostic about the possibility of fine-tunings in both electroweak symmetry breaking and quantities relevant for DM phenomenology,  such as $\Omega_{\chi}^{(\rm th)}$ and $\sigma_{\rm SI, SD}$. We survey  the entire parameter space of thermal and non-thermal neutralino DM, regardless of tuning.  Indeed, it may be reasonable for a tuning of parameters to produce the observed $\Omega$ if environmental selection plays a role in the DM abundance (although a tuning that produces a small $\sigma$ would be more surprising).   In addition, we see in \Fig{fig:TemperOmega} that the relic density curves are steep for a wide range of $M_1$, as is expected from the phenomenon of well-tempering.  In such a situation, where a large fraction of parameter space is highly sensitive to parameters, perhaps one should not be surprised to end up with parameters that appear tuned.  Despite these misgivings about avoiding tuned regions, we do view it as interesting to identify regions of parameter space especially sensitive to parameter choice.   At times we will quantify the amount of tuning in $\Omega_{\chi}^{(\rm th)}$ and $\sigma_{\rm SI,SD}$.  In order to identify DM tunings independently of a possible electroweak tuning, we use a measure, defined in App.~\ref{app:tune}, that mods out the dependence on the direction of parameter space that changes the electroweak VEV.

\section{Suppression of Dark Matter Scattering}  
\label{sec:scattsupp}

In general, SI scattering is mediated by squarks, $Z$ bosons, or Higgses.  Since squark-mediated scattering is model dependent---its effects become negligible for sufficiently heavy squark masses---we postpone our discussion of this scenario to \Sec{subsec:squarks}.   Similarly, we neglect scattering mediated by the heavy Higgses,  which decouple quickly in the limit $m_A \gg m_Z$.
This leaves scattering mediated by the light Higgs or $Z$, which may be suppressed compared to naive expectations by two effects.  First, a suppression results whenever the DM is close to a pure gaugino or Higgsino, and second, the relevant amplitude exactly vanishes at critical values of parameters, which we call blind spots.

\subsection{Suppression from Purity}
\label{subsec:purity}

The leading SI scattering is mediated by Higgs exchange, and the relevant coupling,  $c_{h \chi \chi}$, originates, at tree-level, from gaugino Yukawa couplings of the form $h^\dagger \tilde{h} \tilde{b}$ and $h^\dagger \tilde{h} \tilde{w}$.  Hence SI scattering is suppressed if $\chi$ is dominantly Higgsino or dominantly gaugino.  Similarly, SD scattering does not occur for pure bino or pure wino because they do not couple to the $Z$, and likewise for pure Higgsino because as a Dirac fermion it carries no chiral couplings to the $Z$.

Recall that the neutralino mass matrix takes the form,
\bea
M_\chi &=& \left(
\begin{array}{cccc}
M_1 & 0 &  -\frac{1}{2} {g'v  \cos \beta } & \frac{1}{2} {g' v \sin \beta} \\
0      & M_2 &  \frac{1}{2} {g v  \cos \beta } & -\frac{1}{2} {g v \sin \beta } \\
 -\frac{1}{2} {g' v \cos \beta }   &  \frac{1}{2} {g v \cos \beta }   &  0 & -\mu \\
 \frac{1}{2} {g' v \sin \beta}  & - \frac{1}{2} {g' v \cos \beta }   &   -\mu  & 0.
 \label{eq:neutralinomm}
\end{array}\right).
\eea
Since we are interested in $M_1,M_2, \mu > M_Z$, \Eq{eq:neutralinomm} shows that gaugino-Higgsino mixing is generically small, so that some cross-section suppression is expected for a typical point in parameter space.  However, if the two lightest states are nearly degenerate the mixing between them, $\theta$, can be appreciable, giving $c_{h \chi \chi}, c_{Z \chi \chi} \propto \theta$ with \cite{ArkaniHamed:2006mb} ,
\be
\theta = \frac{(\sin \beta \pm \cos \beta) \sin \theta_W}{\sqrt{2}} \,\, \left( \frac{M_Z}{\Delta M} \right),
\label{eq:thetagh}
\ee
for gaugino/Higgsino DM and
\be
\theta = \frac{\sin 2 \beta \sin 2\theta_W}{2} \,\, \left( \frac{M_Z^2}{\mu (M_2-M_1)} \right),
\label{eq:theta12}
\ee
for bino/wino DM\@.  Both results are valid for a mass splitting $\Delta M > M_Z$; $\theta_W$ is the weak mixing angle and the signs in \Eq{eq:thetagh} refer to the cases $\mu \simeq \pm M_i$.

For successful thermal freeze-out with $\Omega_{\chi}^{(\rm th)} = \Omega_{\rm obs}$ some degree of degeneracy is required, as seen in \Fig{fig:TemperOmega}, so that SI and SD scattering may not be suppressed. However, significant suppression is expected for typical parameters in the cases of non-thermal or multi-component DM.

So far we have been considering tree-level scattering, which vanishes for pure gaugino or Higgsino.  But scattering between a pure Higgsino or wino and nuclei is generated by loop diagrams, for example 1-loop box diagrams with the exchange of two gauge bosons.  Naively the 1-loop scattering has a SI cross-section of $\sigma \sim 10^{-(47-46)}~\mathrm{cm}^2$, which could be probed by the next generation of direct detection experiments.  However, an accidental cancellation~\cite{Hisano:2011cs, Hill:2011be} among various 1 and 2 loop diagrams leads, for pure Higgsino or wino, to cross-sections too small to probe at XENON1T, $\sigma < 10^{-47}~\mathrm{cm}^2$.

\subsection{Suppression from Blind Spots}  
\label{subsec:blindspots}

To obtain a formally vanishing tree-level cross-section through purity, the gauginos or Higgsinos must be completely decoupled, $M_{1,2}$ or $\mu \rightarrow \infty$.  We now consider a different possibility: special choices of parameters where the tree-level cross-section vanishes identically.  At these {\it blind spots}, the gaugino and Higgsino masses are finite and the mixing is non-zero.

Throughout our analysis, we assume that $M_1$, $M_2$, and $\mu$ are real parameters, but carry arbitrary signs.  However, only two of the three apparent signs are physical, as is clear from the field redefinition
\bea
\tilde b &\rightarrow& i \tilde b\\
\tilde w &\rightarrow& i \tilde w \\
\tilde h_{u,d} &\rightarrow& -i \tilde h_{u,d} , 
\eea
which is equivalent to simultaneously sending the all the mass parameters $M_1$, $M_2$, and $\mu$ to minus themselves.  In many of our results, we will eliminate the unphysical, overall sign by fixing the sign of a single theory parameter to be positive.

 \begin{table}[t]
\begin{center}
\begin{tabular}{|c|c|c|}
\hline
 $\mathbf{ m_\chi}$ & {\bf condition} & {\bf signs} \\ 
\hline \hline
$M_1$ & $M_1 + \mu \sin 2 \beta = 0$ & $\mathrm{sign}( M_1 /\mu) = -1$ \\
\hline
$M_2$ & $M_2 + \mu \sin 2 \beta = 0$ & $\mathrm{sign}(M_2 /\mu) = -1$ \\
\hline
$-\mu$ & $\tan \beta = 1$ & $\mathrm{sign}(M_{1,2}/\mu) = - 1^*$ \\
\hline
$M_2$ & $M_1 = M_2$ &  $\mathrm{sign}(M_{1,2} /\mu) = - 1$ \\
\hline
\end{tabular}
\end{center}
\caption{\label{tab:blindspots}
Table of SI blind spots, which occur when the DM coupling to the Higgs vanishes at tree-level.  The first and second columns indicate the DM mass and blind spot condition, respectively.  All blind spots require relative signs among parameters, as emphasized in the third column.  $^*$For the third row, the blind spot requires that $\mu$ and $M_1 ~(M_2)$ have opposite signs when $M_2 ~(M_1)$ is heavy.
}
\end{table}

Let us denote the mass eigenvalues of $M_\chi$ by $m_{\chi_i}(v)$, where $i=1,2,3,4$ and $m_{\chi_1} \equiv m_\chi$ is the DM mass.  Here we have emphasized the explicit $v$ dependence in the masses.
The coupling of  any of neutralino to the Higgs boson can then be obtained by replacing $v\rightarrow v + h$, as dictated by low-energy Higgs theorems \cite{Ellis:1975ap,Shifman:1979eb}:
\bea
{\cal L}_{h \chi\chi} &=& \frac{1}{2} m_{\chi_i}(v+h) \chi_i \chi_i \\
&=&   \frac{1}{2} m_{\chi_i}(v) \chi_i \chi_i+\frac{1}{2} \frac{\partial m_{\chi_i}(v)}{\partial v} h \chi_i \chi_i + {\cal O}(h^2),
\eea
which implies that $\partial m_{\chi_i}(v) / \partial v=c_{h \chi_i \chi_i}$~\cite{Cohen:2011ec,Cheung:2011aa}.  

Consider the characteristic equation satisfied by one of the eigenvalues $m_{\chi_i}(v)$,
\bea
\det (M_\chi - \mathbb{1} m_{\chi_i}(v))&=& 0.
\eea
Differentiating the left-hand side with respect to $v$ and setting $\partial m_{\chi_i}(v) / \partial v=c_{h \chi_i\chi_i} = 0$, one then obtains a new equation which defines when the neutralino of mass $m_{\chi_i(v)}$ has a vanishing coupling to the Higgs boson\footnote{We have checked that Eq.~\ref{eq:cancel} can also be derived using analytical expressions for bilinears of the neutralino diagonalization matrix from Ref.~\cite{Gounaris:2001fx}.}:
\bea
(m_{\chi_i}(v) + \mu \sin 2\beta)\left(m_{\chi_i}(v) -\frac{1}{2}(M_1 + M_2 +\cos  2 \theta_W (M_1 - M_2)) \right) &=&0.
\label{eq:cancel}
\eea
The above equation implies that for regions in which $c_{h \chi_i\chi_i}=0$,  $m_{\chi_i}(v)$ is entirely independent of $v$.  At such cancellation points, $m_{\chi_i}(v) = m_{\chi_i}(0)$, so the neutralino mass is equal to the mass of a pure gaugino or Higgsino state and $m_{\chi_i}(v) = M_1,M_2,-\mu$. 
As long as \Eq{eq:cancel} holds for the LSP mass, $m_{\chi_1}(v)$, then the DM will have a vanishing coupling to the Higgs boson, yielding a SI scattering blind spot.
It is a nontrivial condition that  \Eq{eq:cancel} holds for the LSP, rather than a heavier neutralino, because for some choices of parameters the DM retains a coupling to the Higgs but one of the heavier neutralinos does not.  We have identified these physically irrelevant points and eliminated them from consideration.  The remaining points are the SI scattering blind spots for neutralino DM,
\bea
\begin{array}{c}
\textrm{\bf spin-independent} \\ \textrm{\bf blind spots} \end{array}&:&
\begin{array}{lll}
m_{\chi_1} &=& M_1, M_2, -\mu \textrm{, and }m_{\chi_1}+\mu \sin 2\beta = 0  \\
m_{\chi_1} &=& M_1 = M_2, 
\end{array}
\label{eq:SIblind}
\eea
where in the first line, $m_{\chi_1} = M_1, M_2, -\mu$, depending on whether the LSP becomes pure bino, wino, or Higgsino, respectively, in the $v\rightarrow 0$ limit.  Note that the blind spots in \Eq{eq:SIblind} only appear for certain choices of relative signs.  In the first line, for example, if $m_{\chi_1} = M_1 (M_2)$, then $\mu$ and $M_1$ ($M_2$) must have opposite signs; when $m_{\chi_1} = -\mu$, then $\mu$ must have the opposite sign of $M_1 (M_2)$ when $M_2 (M_1)$ is heavy.   For the second line, the blind spot occurs if $\mu$ and $M_1 = M_2$ have opposite signs. The complete set of conditions required for a SI blind spot are summarized in Table~\ref{tab:blindspots}.

Destructive interference between light and heavy Higgs exchange may also produce cancellations in the SI cross-section~\cite{Altmannshofer:2012ks}, but these are outside the scope of this work.  We consider interference between Higgs and squark exchange in section~\ref{subsec:squarks}.

Next, let us consider SD scattering, which is mediated by $Z$ boson exchange.  The coefficient of the relevant operator vanishes for neutralino DM when
\bea
\begin{array}{c}
\textrm{\bf spin-dependent} \\ \textrm{\bf blind spot}  \end{array} &:& \tan \beta = 1,
\label{eq:SDblind}
\eea
yielding a blind spot for SD direct detection.    The cancellation of the SD $Z$ boson coupling to DM can be understood from symmetry arguments:  when $v_u = v_d$, the DM Lagrangian enjoys an enhanced symmetry under which $u \leftrightarrow d$.  In this limit left-right parity is restored and hence the parity-violating $Z$ coupling which mediates SD scattering will vanish.

So far our discussion of blind spots has been tree-level.  One may wonder how the blind spots change when loop corrections are included.  Loop corrections have not been computed in the full parameter space, but only for the simplifying assumption of pure DM~\cite{Hisano:2011cs, Hill:2011be}, as discussed above.  But our expectation is that the loop corrections are small,  generically resulting in a small shift in the location of the blind spots.  Moreover, at a typical point in parameter space, the mixing angles are small and the multiloop result for pure Higgsino or wino will approximately apply, leading to a cross-section too small to probe in upcoming experiments like XENON1T\@.  Full consideration of loop corrections is beyond the scope of our study, but we estimate the size of these corrections in App.~\ref{app:loop}.

\section{Bino/Higgsino Dark Matter}
\label{sec:bh}

In this section we consider the present and future status of non-thermal, multi-component and thermal bino/Higgsino DM\@. Mixed bino/Higgsino has been studied in a variety of contexts and more recently has been re-examined in light of results from direct detection experiments \cite{Grothaus:2012js, Perelstein:2012qg, Farina:2011bh}.  Here we take a simplified model approach to bino/Higgsino DM, decoupling all superpartners, other than the bino and Higgsinos, and all Higgs-like scalars other than the SM-like state near 125 GeV\@.  Thus DM is described by just three parameters, $(M_1, \mu, \tan \beta)$.  Our analysis applies to this decoupled limit of the MSSM, NMSSM and to Split Supersymmetry.  At the end of the section we consider additional effects that arise when the squarks are not decoupled.  Some effects from non-decoupling of the wino are illustrated in the next section.  For simplicity we remove a physical phase by imposing CP conservation on the neutralino mass parameters, but we study the effect of the physical sign between $\mu$ and $M_1$.  Our convention is to take $\beta$ in the first quadrant and choose $M_1$ positive, allowing both signs of $\mu$.  Our numerical results here, and in \Sec{sec:bw}, use {\tt MicrOMEGAs~2.45} for cross-sections~\cite{Belanger:2008sj} and relic densities~\cite{Belanger:2006is}.

 \begin{figure}[h!]
\begin{center} 
\includegraphics[scale=.5]{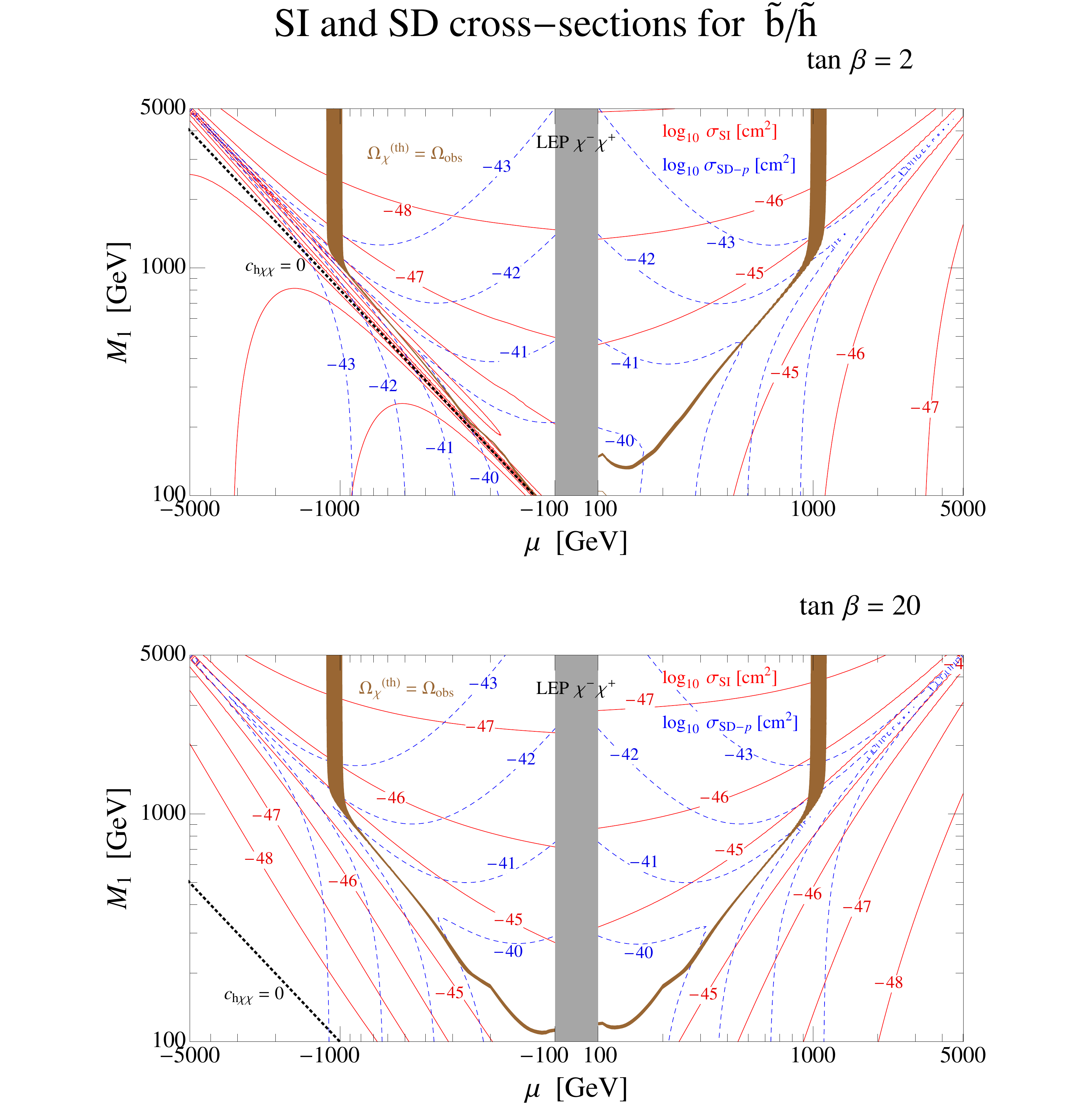}
\end{center}
\caption{ \label{fig:BinoHiggsinoCross}
 Contours of the tree-level cross-sections for SI (solid red) and SD (dashed blue) scattering of bino/Higgsino DM\@. The brown band denotes regions with $\Omega_{\chi}^{(\rm th)}$ within $\pm 3\sigma$ of $\Omega_{\rm obs}$.  The black dashed line is the SI blind spot, $c_{h \chi \chi} =0$, arising from the relation $M_1 + \mu \sin 2 \beta  = 0$.  The central gray region is excluded by LEP.}
\end{figure}

We may understand the results of the following subsections by considering the variation of the SI and SD elastic scattering cross-sections within the bino/Higgsino parameter space, as shown in \Fig{fig:BinoHiggsinoCross}, arising from the tree-level exchange of a SM Higgs of mass 125 GeV for SI and from the $Z$ boson for SD\@.  At large $\tan \beta$ (lower panel) the SI and SD cross-sections become independent of the sign of $\mu$.  This happens because in this regime the bino mixes negligibly with the down-type Higgsino and so the sign of $\mu$ can be removed by a field redefinition and is unphysical. 
  Both the SI and SD contours fall off with increasing $\mu$ or $M_1$ when the other parameter is kept fixed, as expected from the vanishing of the bino/Higgsino mixing angle.   As $\mu$ approaches $M_1$, the mixing angle maximizes, and the contours show a ridge along the line $M_1 \sim \mu$, with the cross-sections dropping off steeply on both sides.  This behavior can be understood from the discussion in section \ref{subsec:purity}; the ridge corresponds to the region with large mixing between the lightest two nearly degenerate states, given by eq. (\ref{eq:thetagh}), from which it follows that the ridge becomes steeper at large masses.   Furthermore, this region of maximal scattering cross-section coincides with the well-tempered line, since large mixing is also necessary to achieve the observed relic abundance.

The black dashed lines show blind spots for SI scattering with $c_{h \chi \chi}=0$, arising from the relation $M_1 + \sin 2 \beta \, \mu = 0$.  For high $\tan \beta$ this occurs in a region where the SI cross-section is highly suppressed by a small mixing angle, but at low $\tan \beta$ (upper panel) the blind spot cuts a gorge in the ridge at negative $\mu$.  As we will see, the proximity of this blind spot to the region with large mixing angles has important implications for the observability of thermal DM, although the rapid variation of the contours implies an enhanced tuning of the cross-section.

\subsection{Non-thermal Dark Matter with $\Omega_{\chi}=\Omega_{\rm obs}$}

 \begin{figure}[h!]
\begin{center} 
\includegraphics[scale=.5]{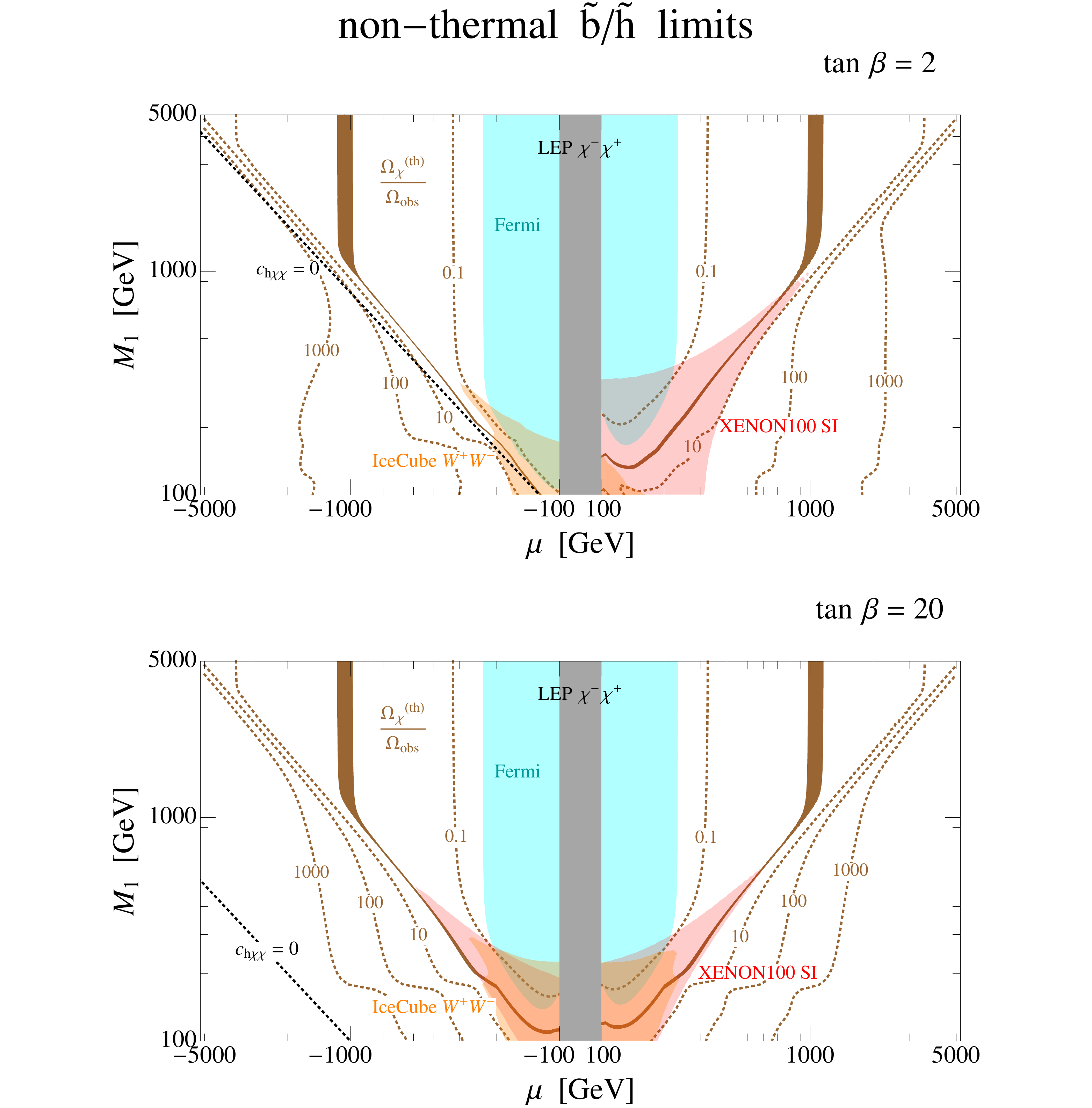}
\end{center}
\caption{\label{fig:BinoHiggsinoLimit}
Current limits on bino/Higgsino DM with $\Omega_{\chi}=\Omega_{\rm obs}$ for $\tan \beta$ = 2 (upper), 20 (lower).  Dotted brown lines are contours of $\Omega_{\chi}^{(\rm th)} / \Omega_{\rm obs}$, and the brown band shows the region having $\Omega_{\chi}^{(\rm th)}$ within $\pm3\sigma$ of $\Omega_{\rm obs}$.   Regions above (below) the brown band require an enhancement (dilution) of the DM abundance after freeze-out.   Regions currently excluded by XENON100, IceCube, Fermi, and LEP are shaded.   The black dashed line is the SI blind spot, $c_{h \chi \chi}=0$, and is close to (far from) the brown band for low (high) $\tan \beta$.}
\end{figure}

We begin by considering the limits on bino/Higgsino DM without imposing that thermal freeze-out provide the observed relic abundance.  \Fig{fig:BinoHiggsinoLimit} depicts contours of $\Omega_{\chi}^{(\rm th)} / \Omega_{\rm obs}$---the ratio of the thermal yield of neutralino DM to the observed relic abundance---together with current experimental constraints in the $(\mu, M_1)$ plane at $\tan \beta = 2,20$.    For $|M_1  | \ll |\mu|$, DM is bino-like and $\Omega_\chi^{(\rm th)}$ is over-abundant; for $|M_1 | \gg |\mu|$, DM is Higgsino-like and  $\Omega_\chi^{(\rm th)}$  is under-abundant.  In these regimes we have evaluated constraints assuming a non-standard cosmological history in which entropy production or non-thermal DM production, respectively, ensures a final neutralino abundance of  $\Omega_{\rm obs}=\Omega_\chi \neq  \Omega_\chi^{(\rm th)}$.

According to \Fig{fig:BinoHiggsinoLimit},  thermal bino/Higgsino DM at low $\tan \beta$ is excluded up to $m_\chi \simeq 800$ GeV for $\mu >0$ but practically unconstrained for $\mu <0$.  At high $\tan \beta$, however, thermal bino/Higgsino DM is excluded for $m_{\chi} \simeq 500$ GeV for either sign of $\mu$.  Meanwhile, non-thermal bino-like or Higgsino-like DM is, at present, rather poorly constrained by direct detection on account of the relatively small mixing, and therefore small couplings to the Higgs and $Z$.  Conversely, indirect detection limits do exclude non-thermal Higgsino-like DM above the $W^+W^-$ threshold up to $\mu \sim 250$ GeV, but well-tempered DM, which has a smaller annihilation cross-section, evades this bound. 

The Fermi limit on DM annihilation to $W^+W^-$ primarily comes from constraints on photons created in the decays of hadrons.  In order to obtain the Fermi exclusion region shown in \Fig{fig:BinoHiggsinoLimit}, we include the annihilation cross-sections to both $W^+W^-$ and $ZZ$, weighted by the relative hadronic branching ratios.
  
\begin{figure}[h!]
\begin{center} 
\includegraphics[scale=.5]{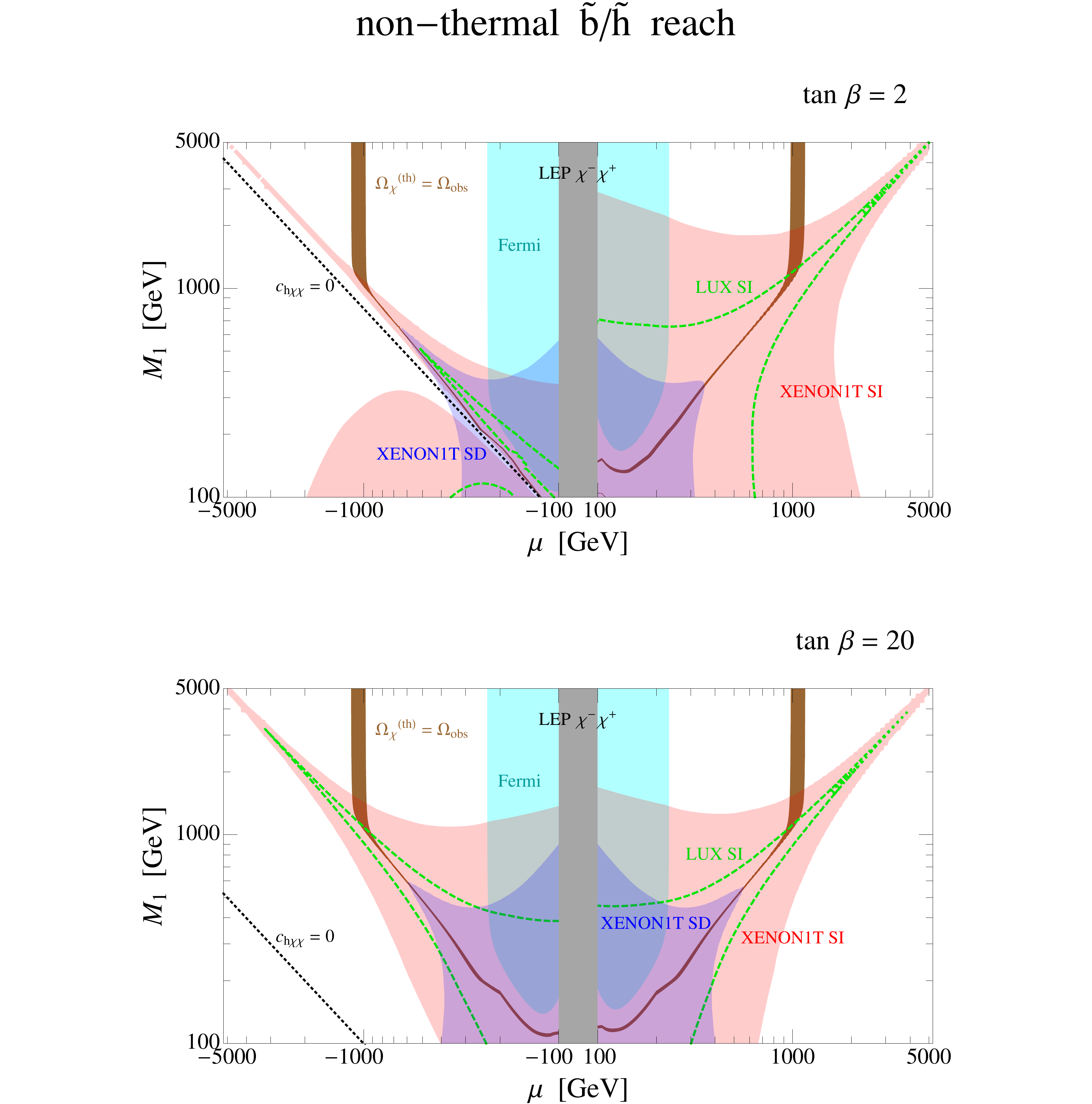}
\end{center}
\caption{\label{fig:BinoHiggsinoReach}
Same as \Fig{fig:BinoHiggsinoLimit} except for future reach rather than current limits.  The dashed green lines show the projected SI reach of LUX, while the shaded regions give the projected reach for XENON1T, both SI and SD\@.  The shaded cyan region is the current Fermi exclusion, as in \Fig{fig:BinoHiggsinoLimit}.}
\end{figure}

 The structure of \Fig{fig:BinoHiggsinoLimit} follows from the fact that the leading experimental constraint on bino/Higgsino DM is on SI scattering at XENON100.  In general, thermal neutralino DM tends to be the most constrained by SI direct detection, simply because DM carries an $\OO(1)$ fraction of bino and Higgsino that furnishes a non-vanishing coupling to the Higgs.  In contrast, parameter regions corresponding to non-thermal DM are more weakly constrained, since pure bino and pure Higgsino DM do not couple directly the Higgs boson.  That said, even well mixed neutralino DM can be decoupled from the Higgs if the theory parameters reside on the SI blind spot defined in \Eq{eq:SIblind},
\bea
c_{h \chi\chi} &\propto& M_1 + \mu \sin 2\beta = 0,
\label{eq:binoblind}
\eea
which is only allowed for $\mu < 0$.   At low $\tan \beta$, the SI blind spot occurs near the well-tempering region, $|M_1| \simeq |\mu|$.   The $\tan \beta =20$ plot in \Fig{fig:BinoHiggsinoLimit} is approximately symmetric under $\mu \leftrightarrow -\mu$, for the reasons noted earlier.

At present, limits on the SD scattering cross-section are dominated by IceCube bounds.  These provide a complementary constraint for lighter DM near the SI blind spots, since SD scattering cross-sections are unaffected by the vanishing of the coupling to the Higgs.  As discussed in \Sec{sec:obs}, IceCube provides bounds on DM annihilation to $W^+W^-$; in order to generate the exclusion regions in \Fig{fig:BinoHiggsinoLimit} and \Fig{fig:BinoHiggsinoReach}, we compare these bounds to the DM annihilation cross-section into $W^+W^-$, $Z h$, and $ZZ$, weighted by their branching ratio to neutrinos relative to $W^+W^-$.  Previous studies~\cite{Kozaczuk:2012vx,Silverwood:2012tp} have also considered limits on neutralino DM annihilation from IceCube, using older data or projections.

\Fig{fig:BinoHiggsinoReach} is identical to \Fig{fig:BinoHiggsinoLimit} except it depicts projected reach instead of current limits.  Comparing \Fig{fig:BinoHiggsinoReach} and \Fig{fig:BinoHiggsinoLimit}, LUX and XENON1T will provide a very powerful probe of both thermal and non-thermal bino/Higgsino DM\@.  Currently only narrow wedges of the $(\mu, M_1)$ plane are excluded.  These wedges lie along the thermal band, but even the exclusion of some thermal regions is marginal and subject to astrophysical uncertainties.  Over the next few years, LUX and XENON1T will explore most of the parameter space with DM masses up to 1 TeV, and much of the region up to 2 TeV, offering a remarkable opportunity for discovery.  If no signal is seen, LUX will exclude a large fraction of thermal bino/Higgsino DM and, XENON1T will exclude the entire parameter space of thermal bino/Higgsino DM for $\tan \beta >2$, except for the case of almost pure Higgsino.   Interesting blind spot regions remain for lower $\tan \beta$, as discussed in the following subsections.

On the other hand, even in the absence of a signal, significant parameter regions for the non-thermal case will remain.   Bino-like DM is permitted for $\mu < 0$ near the SI blind spot for bino-like DM defined in \Eq{eq:binoblind}.   Meanwhile, non-thermal Higgsino-like DM is highly unconstrained at low $\tan \beta$  because it corresponds to the SI blind spot for Higgsino-like DM in \Eq{eq:SIblind},
\bea
c_{h \chi\chi} &\propto& -1 + \sin 2\beta = 0.
\label{eq:Higgsinoblind}
\eea
Some of these allowed regions  will be probed by experiments sensitive to the SD scattering cross-section.  Intriguingly, the case of non-thermal Higgsino DM at low $\tan \beta$ resides simultaneously in a blind spot for SI and SD scattering!  Furthermore, this region allows low values of $\mu$, and therefore relatively natural theories of electroweak symmetry breaking.    In addition, large unnatural regions with $\mu >$  1 - 2 TeV will remain viable, but require late entropy production, especially for low $M_1$.

\subsection{Multi-Component Dark Matter with $\Omega_\chi = \Omega_\chi^{(\rm th)} \leq \Omega_{\rm obs}$}

\begin{figure}[t!]
\begin{center} 
\includegraphics[scale=.5]{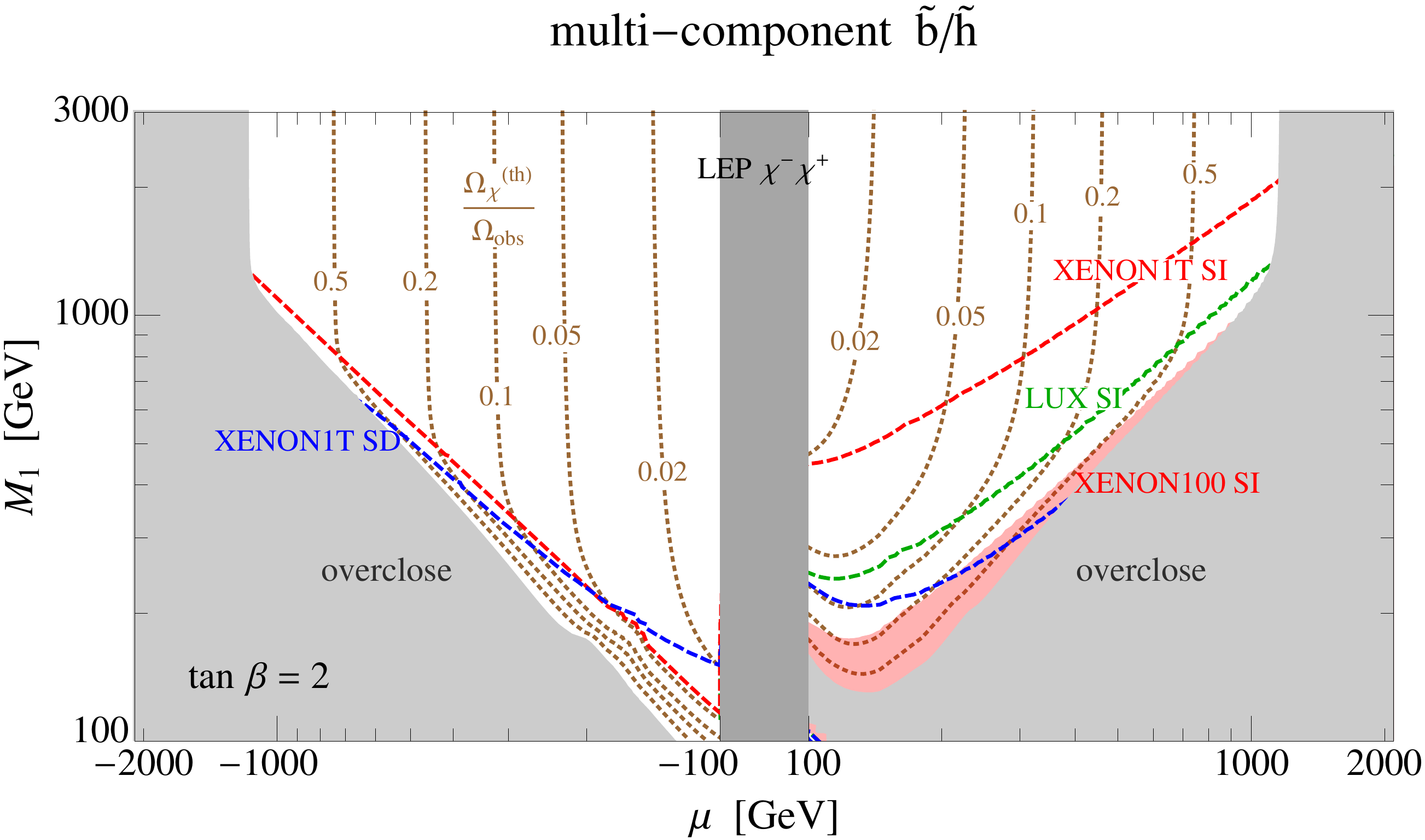}
\end{center}
\caption{\label{fig:multi} 
Limits and projected reaches for multi-component bino/Higgsino DM with $\Omega_\chi = \Omega_\chi^{(\rm th)}$.  Dotted brown lines are contours of $\Omega_{\chi}^{(\rm th)} / \Omega_{\rm obs}$ for $\tan \beta$ = 2.  The light gray regions are excluded by overabundance of neutralino DM, while the edge of this region has $\Omega_\chi^{(\rm th)} = \Omega_{\rm obs}$.  In the remainder of the plane $\chi$ is just one component of multi-component DM\@.  The present limit from XENON100 is shown shaded, while the projected reaches of LUX and XENON1T, both SI and SD, are shown as dashed lines.}
\end{figure}
 
Here we repeat the analysis of the previous section under the assumption that the present day relic abundance of neutralino DM is given by $\Omega_\chi = \Omega_\chi^{(\rm th)}$, with the balance of cosmological DM arising from some other source.  \Fig{fig:multi} depicts both the current limits and the projected reach for such multi-component neutralino DM, for $\tan \beta = 2$.  Region shaded light gray have $\Omega_\chi^{(\rm th)} / \Omega_{\rm obs} > 1 $ and are thus excluded, while regions with $\Omega_\chi^{(\rm th)} / \Omega_{\rm obs} < 1 $ have a depleted abundance of neutralino DM\@.  Direct detection limits are then ameliorated, since the rate of WIMP-nucleon scattering is proportional to the incident flux of DM particles, and thus to $\Omega_\chi^{(\rm th)} / \Omega_{\rm obs}$.  DM annihilation cross-sections are suppressed by the square of this ratio, making indirect detection limits irrelevant.  The edge of the light gray shaded region has $\Omega_\chi^{(\rm th)} =\Omega_{\rm obs}$ and therefore has thermal bino/Higgsino DM, which is explored for all $\tan \beta$ in the next sub-section.

It would appear to require a coincidence for otherwise unrelated stable particles to have comparable relic abundances, as would be needed for multi-component neutralino DM with an $\mathcal{O}(1)$ fraction of $\Omega_{\rm obs}$.  However, environmental selection may provide a possible explanation \cite{Tegmark:2005dy,Hall:2011jd}.  While the expected abundances in the various components depend on multiverse distribution functions, it is likely that they are very roughly comparable.  Alternatively, multiple sectors of a theory may participate in the WIMP miracle, in which case each WIMP would independently attain a thermal abundance near $\Omega_{\rm obs}$.   Consider for example the case that the bino/Higgsino makes up a third of the total DM\@.  Current limits from XENON100 and IceCube allow any mass above the LEP exclusion limit.  However, XENON1T will probe a large and interesting region, pushing the mass up to about 500 GeV if no signal is seen.  Much of the remaining space is dominantly Higgsino.

\subsection{Thermal Dark Matter with $\Omega_\chi = \Omega_\chi^{(\rm th)} =\Omega_{\rm obs}$}

\begin{figure}[h!]
\begin{center} 
\includegraphics[scale=.33]{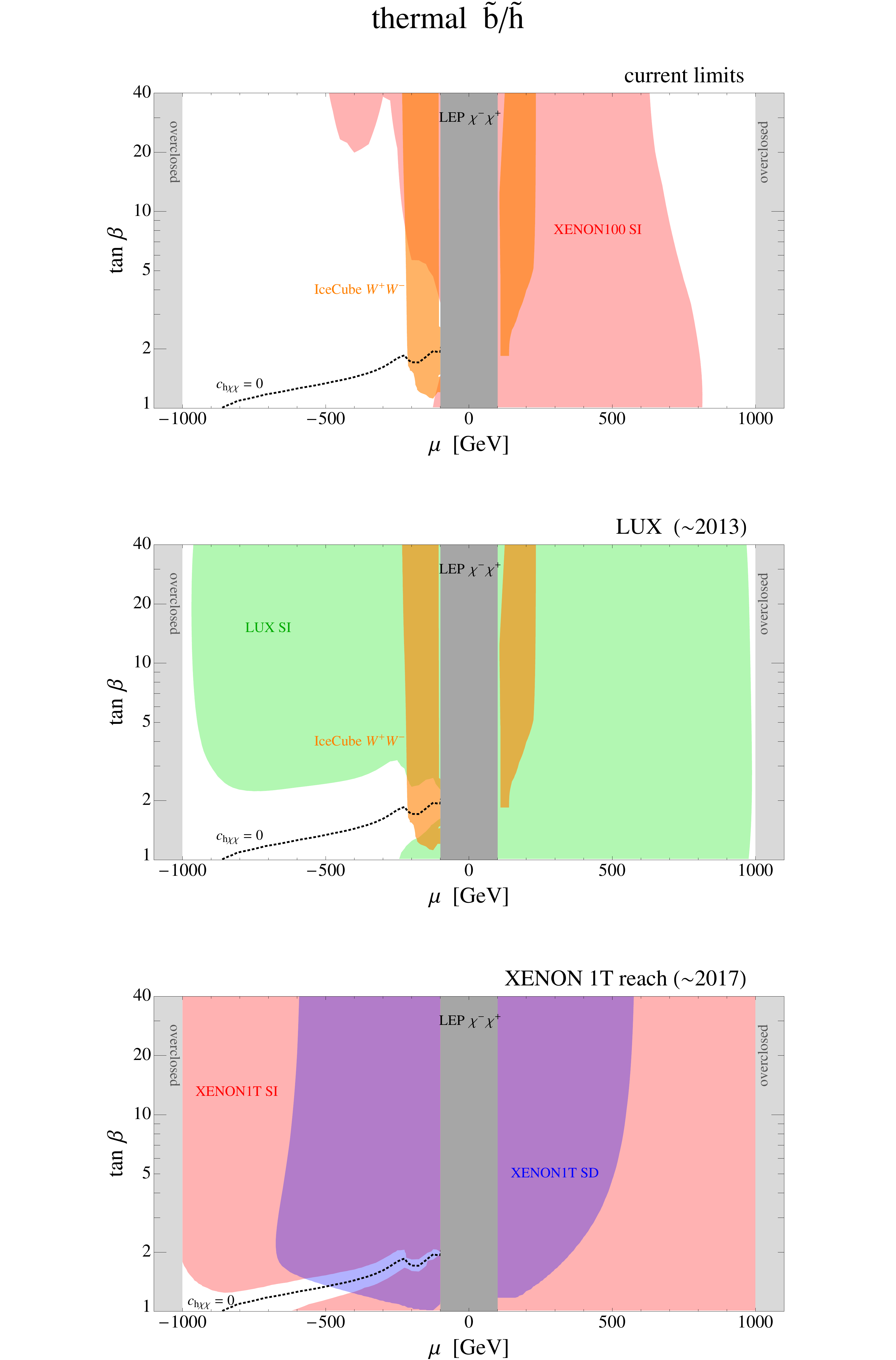}
\end{center}
\caption{\label{fig:BH_WellTemper}
Current limits from XENON100 and IceCube (top), expected reach of LUX and IceCube (middle), and expected reach of XENON1T (bottom) for SI and SD scattering, fixing $M_1$ at every point to accommodate $\Omega_{\chi}^{(\rm th)} =\Omega_{\rm obs}$.  
The black dashed line is the blindspot, $c_{h \chi \chi}=0$.  }
\end{figure}

We now restrict our analysis to well-tempered bino/Higgsino DM\@.  We fix the value of $M_1$ using the relic abundance constraint, $\Omega_\chi^{(\rm th)} = \Omega_{\rm obs}$, reducing the parameter space by one dimension.  Thus we can show the entire parameter space relevant for thermal bino/Higgsino DM in the $(\mu,\tan\beta)$ plane.  \Fig{fig:BH_WellTemper} depicts current limits and projected reach for thermal neutralino DM as a function of these parameters.  For $\mu > 0$, SI direct detection currently rules out thermal bino/Higgsino DM for $\mu \lesssim 650-800$ GeV, depending on $\tan\beta$.  $\mu < 0$, however, is almost completely unconstrained, except for a small region around the $t\bar{t}$ threshold for DM annihilation at high $\tan\beta$.  

Future direct detection experiments will cover the entire well-tempered parameter space for $\mu > 0$, and almost all of it for $\mu < 0$, with the exception of a region around the blind spot cancellation given by \Eq{eq:binoblind}.  The DM coupling to the $Z$ does not vanish at the SI blind spot, however, so that SD direct detection will set complementary limits in this region, with XENON1T probing up to $\mu \sim -500$ GeV.

\begin{figure}[t!]
\begin{center} 
\includegraphics[scale=.425]{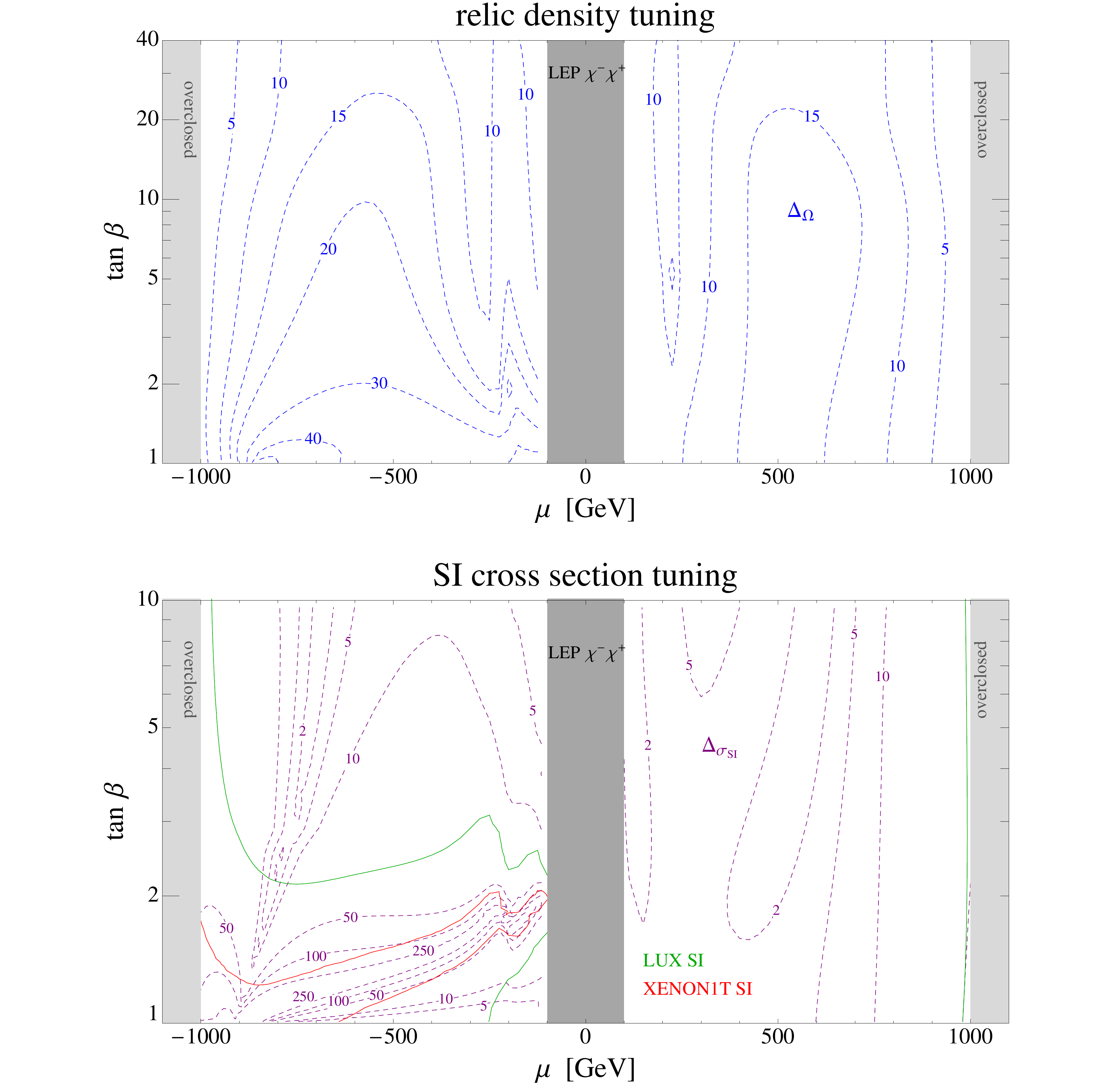}
\end{center}
\caption{\label{fig:BH_Tuning}
Relic density and SI cross-section tuning for well-tempered bino/Higgsino, with the reach of LUX and XENON1T also shown in the second panel.  Tuning of the relic density is typically between 2 - 10\%.  If XENON1T does not see a signal, tuning of the SI cross-section will be $\lesssim$ 1\%.  The interpretation of this as being unnatural is unclear however, as most of the region of $1< \tan \beta <2$ with $\mu <0$ has large $\Delta_\sigma$.  We describe our methodology for computing tuning in App.~\ref{app:tune}.}
\end{figure}

\Fig{fig:BH_Tuning} quantifies the fine-tuning of thermal bino/Higgsino DM\@.  These plots depict the sensitivity of the relic density and the SI scattering cross-section with respect to the ultraviolet parameters in the theory---namely, $M_1, M_2, \mu ,m_{H_u}^2, m_{H_d}^2, B\mu$ defined at the weak scale---using the measure defined in \App{app:tune}.  As might be expected from the steepness of the $\Omega_{\chi}^{\rm (th)}$ curve in \Fig{fig:TemperOmega}, well-tempering requires $\mathcal{O}(10\%)$ tuning throughout the parameter space.  A similar, relatively mild tuning of the direct detection cross-section is required to evade the upcoming LUX bounds on SI scattering.  Evading the limits from XENON1T, however,  requires a cross-section tuning of at least $1\%$, suggesting that it would be unlikely for thermal bino/Higgsino DM to remain hidden.

This tuning of the cross-section, however, is directly correlated with that required to obtain the correct relic abundance at low $\tan\beta$, since both a large mixing angle and a small Higgs coupling require $|\mu| \simeq |M_1|$.  Furthermore, many theories, both natural and unnatural, require small $\tan\beta$ in order to explain the 125 GeV Higgs mass, as in $\lambda$SUSY and Split Supersymmetry.  Thus the region of parameter space which evades XENON1T is exactly the same region in which one might expect to find oneself given both the Higgs mass and the observed relic abundance.  In this case, perhaps a large tuning of the cross-section should be unsurprising, since such a tuning is generic within the allowed parameter space.

\subsection{Squark Effects}
\label{subsec:squarks}
Our analysis thus far has neglected the effects of squarks.  For the natural theories, however, it is reasonable to consider squark masses and $\mu$ that are not exceedingly large, so as not to exacerbate fine-tuning in electroweak symmetry breaking.  In many such scenarios, the DM is mixed bino/Higgsino, and the effects of light squarks can play an important role on the physics~\cite{Drees:1993bu, Feng:2010ef, Hisano:2011um}.

\begin{figure}[h!]
\begin{center} 
\includegraphics[scale=.55]{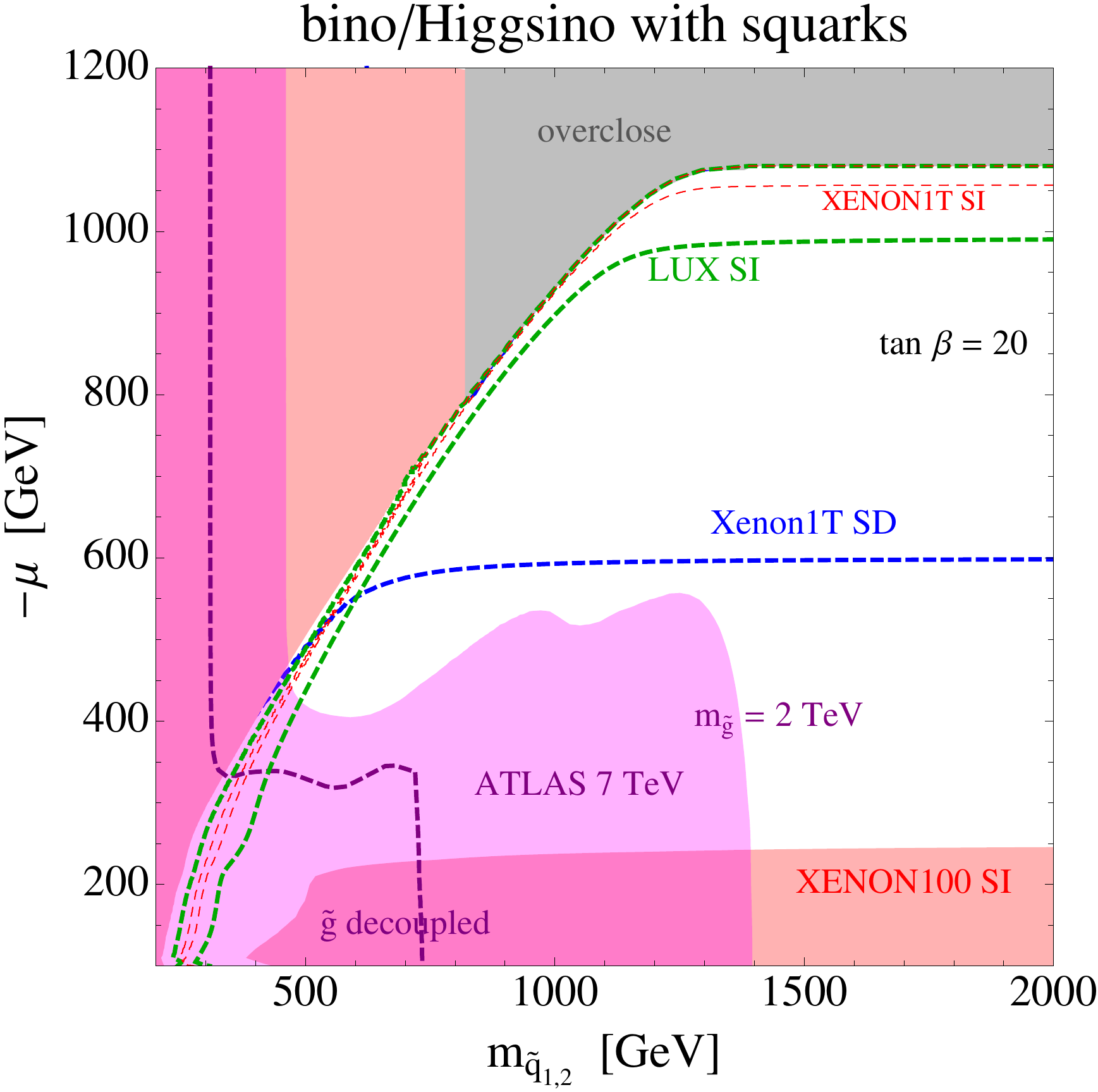}
\end{center}
\caption{\label{fig:BH_Squark}
Impact of squarks on thermal bino/Higgsino DM, with $\mu<0$ and $\tan \beta = 20$.  At each point $M_1$ has been chosen so that $\Omega_{\chi}^{(\rm th)} = \Omega_{\rm obs}$, except in the gray region where freeze-out always yields overclosure.  The upper left region, where freeze-out is dominated by squark-neutralino coannihilation, is excluded by XENON100.  However, in the lower right region the XENON100 limit becomes less powerful as the s-channel squark exchange amplitude has the opposite sign to the t-channel Higgs exchange diagram.  The purple region is excluded by an LHC search for jets and missing transverse energy, with the gluino mass fixed at 2 TeV\@.  This ATLAS search becomes less powerful as the gluino mass is increased, and the excluded region becomes bounded by the purple dashed line if the gluino is decoupled.  The currently allowed region, shown in white, mostly has a SI scattering cross-section that is not far below the current bound, so that LUX will have a large discovery potential.  In the absence of a signal at LUX (XENON1T) the only surviving region will be the narrow band between the dashed green (red) lines. }
\end{figure}

\Fig{fig:BH_Squark} depicts present limits and future reach for bino/Higgsino DM including the effects of light squarks at $\tan \beta = 20$ and $\mu < 0$.  The analogous constraints for $\mu >0$ are more stringent.  Here we have chosen degenerate 1st and 2nd generation squarks at $m_{\tilde q}^2$, with  the 3rd generation decoupled for simplicity.  Squarks of the first generation tend to have the biggest effect because their exchange allows the bino to couple directly to valence quarks.   At each point in the plot we have fixed $M_1$ to accommodate thermal DM, $\Omega_{\chi}^{(\rm th)} = \Omega_{\rm obs}$.  In the limit of $m_{\tilde q}^2 \rightarrow \infty$, this plot asymptotes to the current XENON100 limit  shown for bino/Higgsino DM in \Fig{fig:BH_WellTemper}.

The bino/Higgsino/squark space of \Fig{fig:BH_Squark} can be divided into two regions, depending on whether the squarks are heavier, or lighter than the Higgsino.  When the squarks are lighter than the Higgsino, $m_{\tilde q} < |\mu|$, the correct abundance follows from squark/bino coannihilation.  We find that this entire region of \Fig{fig:BH_Squark} is ruled out by XENON100.  When $m_{\tilde q} > |\mu|$, the abundance follows from bino/Higgsino well-tempering, as in Figs.~\ref{fig:TemperOmega} and~\ref{fig:BH_WellTemper}.  In this region, we see that as the squark masses drop the XENON100 limit disappears.  There are two important effects contributing to this, (1) the contribution to the SI scattering amplitude from $s$-channel squark exchange destructively interferes with the contribution from $t$-channel Higgs boson exchange, and (2) squark diagrams increase the annihilation rate, leading to a smaller bino-Higgsino mixing angle and a smaller DM-DM-Higgs coupling, when the relic density is fixed.

Interestingly, as the squark mass approaches the LSP mass, limits from supersymmetry searches at the LHC are also alleviated.  In particular, we have plotted the limit on the squark/bino simplified model of \cite{:2012rz}.  The purple dashed line represents the limit, presented by ATLAS, when the gluino is decoupled.  The purple shaded region represents our estimate of the limit, when $m_{\tilde g} = 2$ TeV, which is stronger due enhanced squark production with a $t$-channel gluino.  In order to estimate this limit, we have reproduced the ATLAS search using  {\tt Pythia 6.4}~ \cite{Sjostrand:2006za} for event generation, {\tt PGS} for crude detector simulation~\cite{pgs} and {\tt NLLfast} for the NLO~\cite{Beenakker:1996ch} and NLL~\cite{Beenakker:2009ha} squark cross-sections.  Both constraints are somewhat weakened near the region of bino/squark degeneracy.  Regarding future reach, \Fig{fig:BH_Squark} indicates that the projected SI and SD constraints from XENON and LUX rule out a large fraction of the allowed space of bino/Higgsino DM with a light squark.

\section{Bino/Wino(/Higgsino) Dark Matter}
\label{sec:bw}

We now consider the effects of including the wino in the spectrum.  Compared to the previous section, reintroducing the wino adds an extra parameter, so that now we have  a four dimensional parameter space of $(M_1, M_2, \mu, \tan \beta)$.  In general, the LSP is now a combination of bino, Higgsino and wino, but much of our attention will focus on the case of a dominant bino/wino mixture.  Even when the dark matter has a very small Higgsino component, the value of the $\mu$ parameter is crucial for direct detection: in the limit of decoupled $\mu$, bino/wino dark matter has vanishingly small couplings to the Higgs and $Z$ bosons.  While mixed bino/wino dark matter has been explored in the past \cite{BirkedalHansen:2001is,BirkedalHansen:2002am,Baer:2005zc,Baer:2005jq,ArkaniHamed:2006mb,Hisano:2012wm}, it has received substantially less attention than other limits of neutralino dark matter.  Part of the reason for this neglect is theoretical prejudice.  In particular, since thermal bino/wino dark matter originates via coannihilation, working models typically require $M_1 \simeq M_2$, which is disfavored by gaugino unification.  Moreover, as discussed in \Sec{sec:relic}, the coannihilation region is exponentially sensitive to the mass splittings in the theory.  Obviously, non-thermal or multi-component bino/wino dark matter require no such constraint on the masses, and have greater freedom to evade bounds.

\begin{figure}[t!]
\begin{center}
\includegraphics[scale=.525]{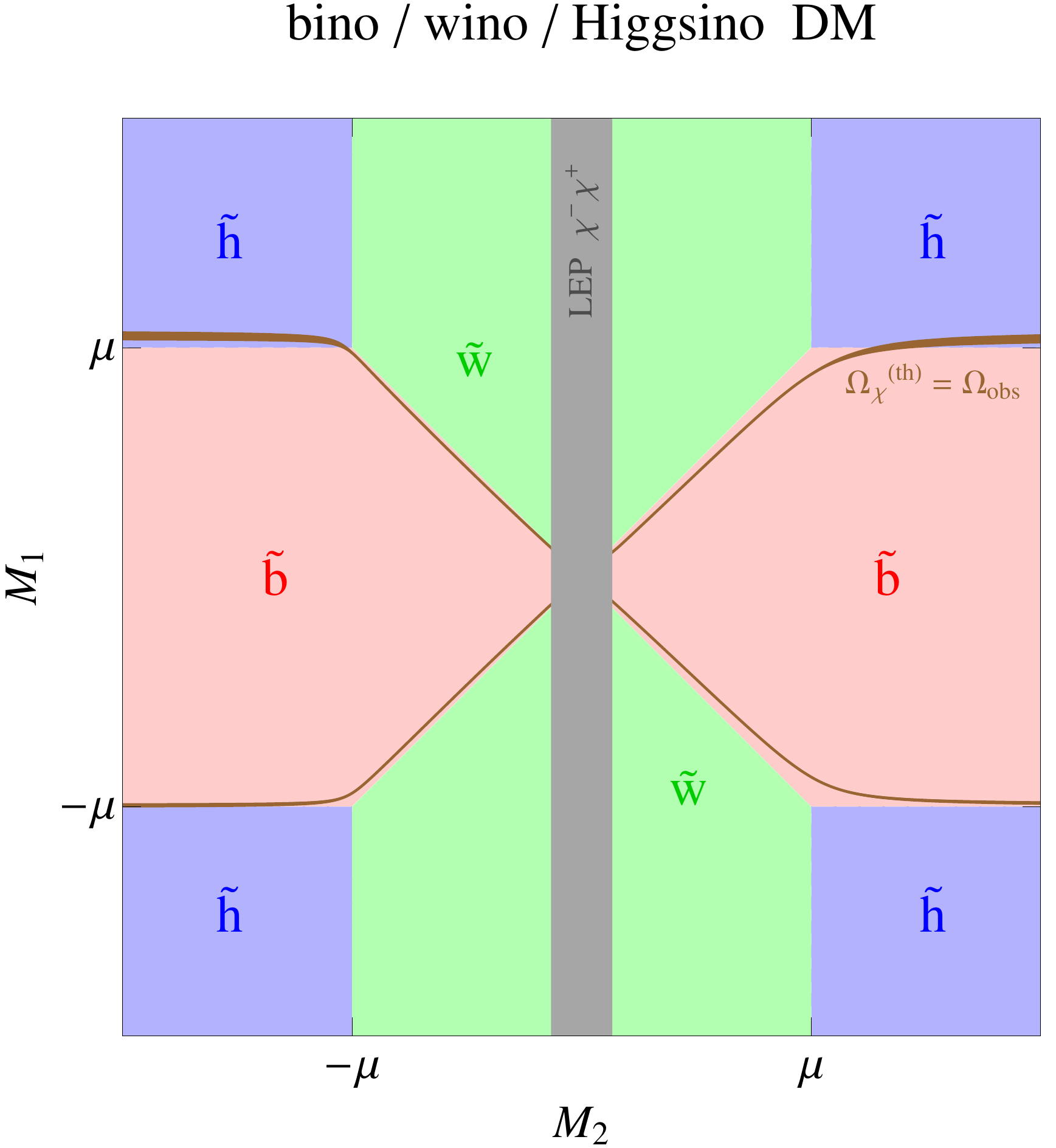}
\end{center}
\caption{\label{fig:BinoWinoScheme}
Phase diagram of neutralino DM in the $(M_1, M_2)$ plane, keeping $\mu$ fixed and less than 1~TeV\@.  The red, green, and blue regions correspond to DM that is mostly bino, wino, or Higgsino-like.  The thermal abundance, $\Omega_\chi^{(\mathrm{th})}$, equals the observed abundance, $\Omega_{\mathrm{obs}}$, along the brown curve, which resides at the boundary of the bino region and wino/Higgsino regions.  Within the bino-like region, the thermal abundance is too large and dilution is required; within the wino and Higgsino regions the thermal abundance is too small and additional neutralino production is required.}
\end{figure}

In this section we present a detailed study of non-thermal, multi-component, and thermal bino/wino DM, focusing on present limits and future reach.  Once again, we remove the physical phases in the neutralino parameters by assuming CP conservation.  In contrast with the previous section, however, there are now two physical, relative signs in the theory.  We continue to take $\beta$ in the first quadrant, and we fix $\mu > 0$ for non-thermal and multi-component DM\@.  For thermal bino/wino DM, however, we use the constraint on relic density to fix $M_1$, which we take to be positive, allowing $\mu$ and $M_2$ to have either sign.

\Fig{fig:BinoWinoScheme} shows a schematic slice of the parameter space relevant for bino/wino DM at fixed $\mu$ and $\tan\beta$.  For $|M_{1,2}| < |\mu|$, the dark matter is dominantly either bino-like or wino-like depending on the hierarchy between $|M_1|$ and $|M_2|$.  Even along the well-tempered line, dark matter is mostly bino, since the relic abundance is set by coannihilation with small mixing angles.  As $|M_1|$ approaches $|\mu|$ at large $|M_2|$, the bino(wino)/Higgsino mixing angle increases until we recover the well-tempered DM considered in the previous section.  In the limit of large $|M_1|$, we may have a mixed wino/Higgsino LSP for $|M_2| \simeq |\mu|$, but the observed relic abundance cannot be achieved for $|\mu|, |M_2| \lesssim 1$ TeV due to the large annihilation cross-section to $W^+W^-$.  The final possibility, also considered in previous sections, is a dominantly Higgsino LSP if $|\mu| < |M_{1,2}|$, which has an over-abundant (under-abundant) thermal relic density for $|\mu| > 1$ TeV $(|\mu| < 1\text{ TeV})$.  In this section we will focus on the upper and lower triangles of \Fig{fig:BinoWinoScheme}, containing non-thermal or multi-component wino DM and bounded by thermal bino/wino DM.

\subsection{Non-thermal Dark Matter with $\Omega_{\chi}=\Omega_{\rm obs}$}

\begin{figure}[t!]
\begin{center} 
\hbox{ \hspace{-1.2cm}
\includegraphics[scale=.40]{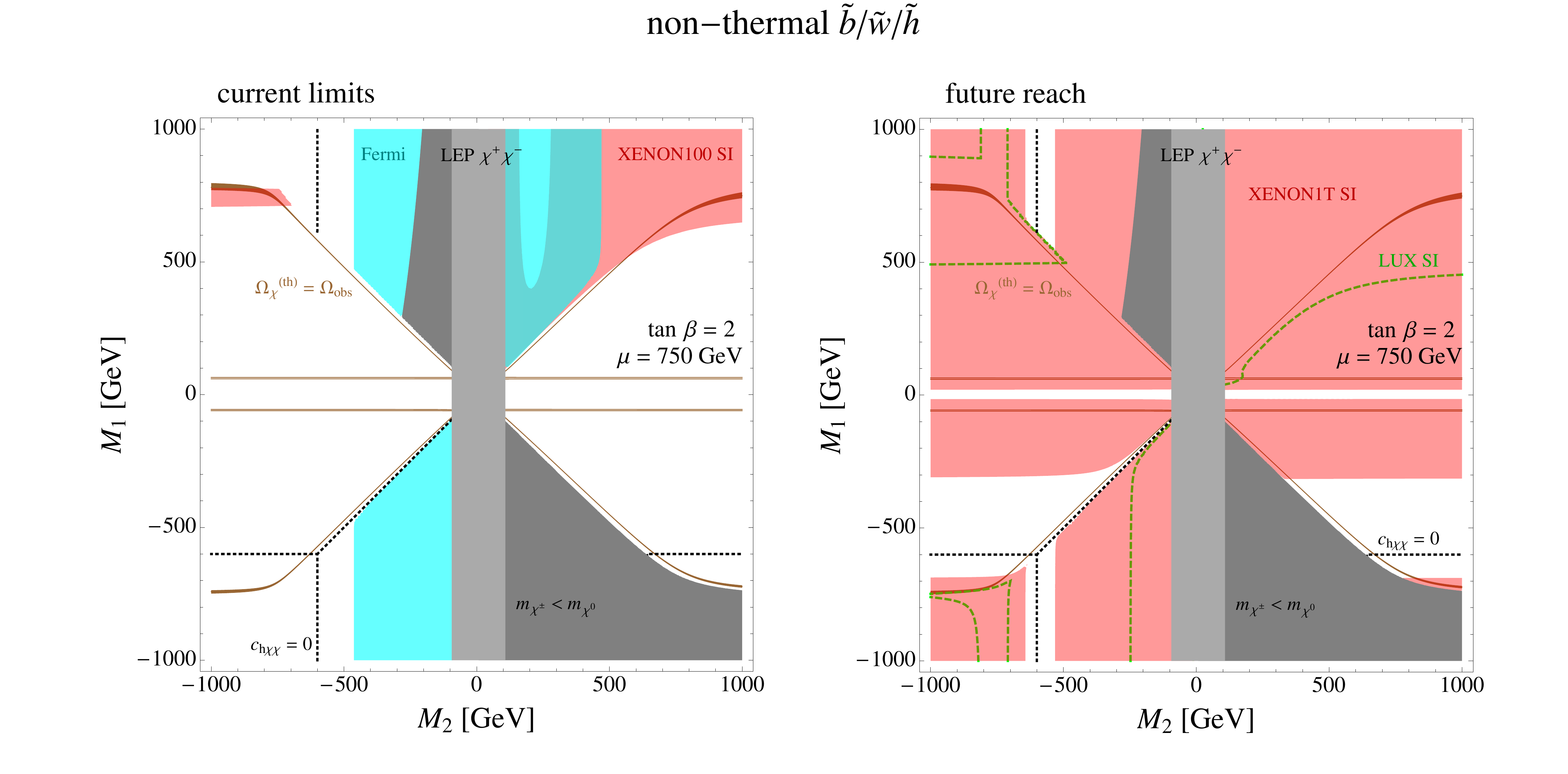} }
\end{center}
\caption{\label{fig:BinoWinoLimit} 
Current limits and future reach for dark matter with $\Omega_{\chi}=\Omega_{\rm obs}$.  The brown band shows the region having $\Omega_{\chi}^{(\rm th)}$ within $3\sigma$ of $\Omega_{\rm obs}$.   Regions with $|M_1|$ larger (smaller) than the brown band require an enhancement (dilution) of the dark matter abundance after freeze-out.   Regions currently excluded by XENON100 and FERMI are shaded in the left-hand figure, and the projected SI reach is depicted with dashed green lines for LUX and in shaded red for XENON1T on the right.  The horizontal (vertical) black dashed lines are the SI blind spot, $c_{h \chi \chi}=0$, for bino-like (wino-like) dark matter.}
\end{figure}

\Fig{fig:BinoWinoLimit} shows the current limits and future reach for neutralino dark matter in a slice of the full parameter space, in particular in the $(M_2, M_1)$ plane at $\tan \beta = 2$ and $\mu = 750$ GeV\@.   The narrow brown bands in these figures show regions with $\Omega_{\chi}^{(\rm th)}$ within $3\sigma$ of $\Omega_{\rm obs}$.   The four physically independent sign choices are realized by allowing $M_{1,2}$ to range over both positive and negative values, yielding four quadrants.  The relic thermal abundance depends largely on gauge interactions, which are independent of the sign choice, so the brown bands are largely the same in the four quadrants.  The horizontal brown bands occur at $|M_1| = m_h/2$ correspond to annihilation of mainly bino dark matter through the Higgs pole.  The brown bands with slope near $45^\circ$ have $|M_1| \sim |M_2| <  \mu$ correspond to bino/wino dark matter, and are narrow because coannihilation is operative.  As $|M_2|$ increase these bands flatten out and represent bino/Higgsino dark matter as the wino decouples.  Here the bands are thicker as coannihilation is no longer present.  The nature of the LSP varies across the $(M_2, M_1)$ plane of \Fig{fig:BinoWinoLimit} as described in the beginning of this section.  At low $|M_2|$ there is a light chargino that is excluded by LEP, as shown by the light gray band.  The dark gray area is also excluded because the LSP is a chargino.

The dark matter direct detection limits and reaches of \Fig{fig:BinoWinoLimit} assume that $\Omega_{\chi}=\Omega_{\rm obs}$ throughout the plane. This requires that, for any $M_2$, regions with $|M_1|$ larger (smaller) than that giving the brown bino/wino or bino/Higgsino band requires enhancement (dilution) of the dark matter abundance after freeze-out.  The shaded red region of \Fig{fig:BinoWinoLimit} is excluded by the recent XENON100 search for SI scattering of dark matter.  This region is highly asymmetric in the four quadrants because direct detection depends on the Higgs coupling $c_{h \chi \chi}$ which depends critically on the sign choices.   In particular, the blind spots for SI scattering of \Eq{eq:SIblind}, with $c_{h \chi \chi}=0 $, occur when $M_1 = - \mu \sin 2\beta < |M_2|$ and $M_2 = - \mu \sin 2\beta < |M_1|$, as shown by the horizontal and vertical dashed black lines in \Fig{fig:BinoWinoLimit}, respectively.  A third blind spot occurs with $0 > M_1 = M_2> - \mu \sin 2 \beta$.   It is only the quadrant which has no blind spot ($M_1, M_2 > 0$) that is substantially constrained by current direct detection bounds from XENON100, being largely excluded except for a small region around the $t\bar{t}$ threshold.   On the other hand, the remaining sign combinations of $M_1$ and $M_2$ are more or less unconstrained.

Thermal dark matter is excluded for $m_\chi \geq 500$ GeV for $M_1,M_2 > 0$.  In contrast with the limits from bino/Higgsino dark matter, limits on bino/wino DM actually become more stringent at larger values of $M_1$ and $M_2$.  This behavior arises simply because SI scattering from Higgs boson exchange for bino/wino DM is mediated by the Higgsino fraction of the LSP\@.  As $M_1$ and $M_2$ increase, then at fixed $\mu$, the dark matter acquires a larger Higgsino fraction and SI scattering is proportionally larger.
Present limits on thermal bino/wino DM are weak because $\mu$ can be modestly decoupled while still accommodating a thermal relic abundance.  This is possible because the dominant process during thermal freeze-out is coannihilation of bino-like dark matter with charged and neutral winos.  Even for larger values of $\mu$ than shown in \Fig{fig:BinoWinoLimit}, the dark matter can be thermally equilibrated with the winos,  permitting efficient coannihilation.

\begin{figure}[t!]
\begin{center} 
\includegraphics[scale=.38]{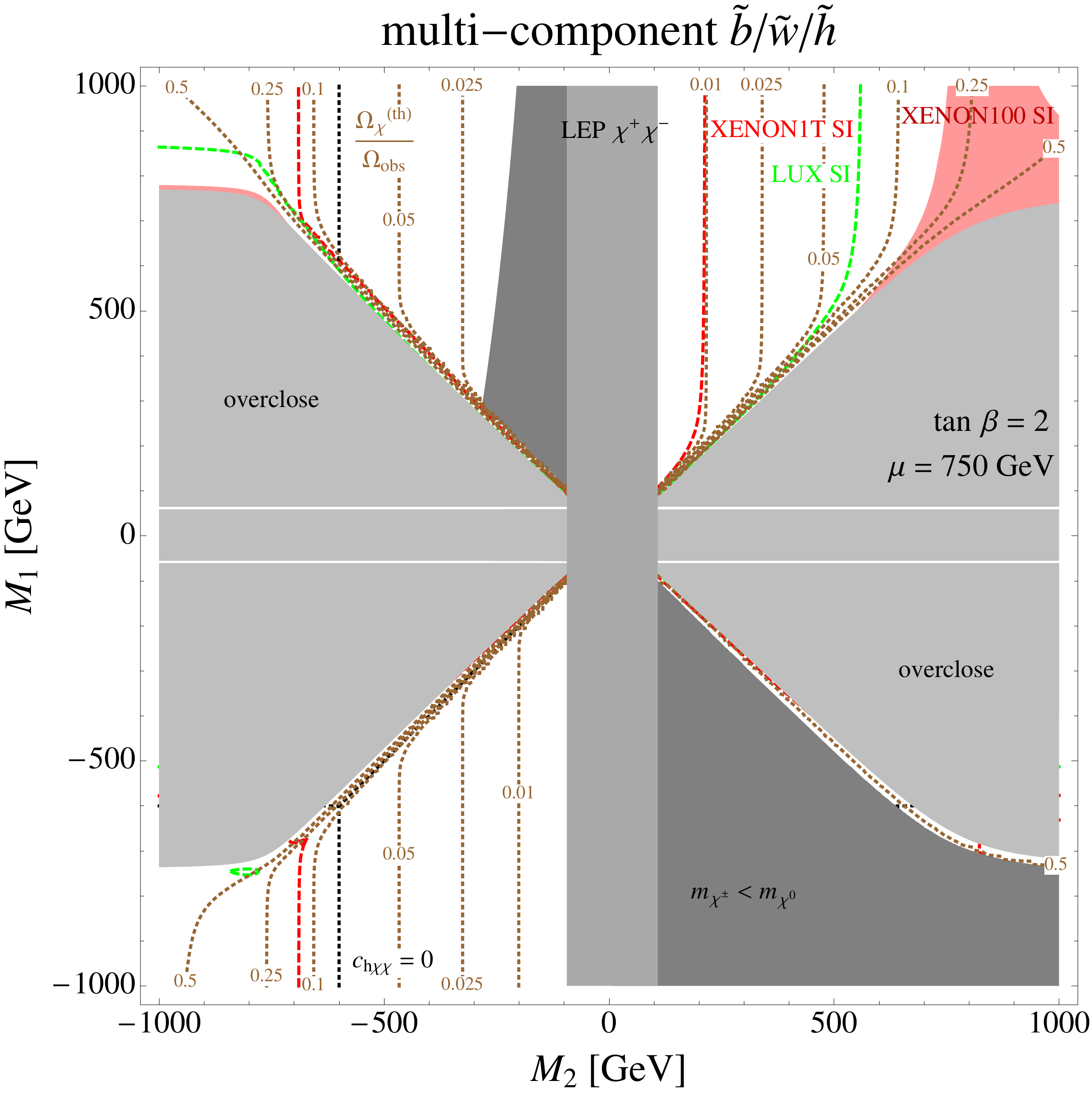}
\end{center}
\caption{\label{fig:BinoWinoSubdominant}
Limits and projected reaches for multi-component dark matter with $\Omega_\chi = \Omega_\chi^{(\rm th)}$ for $\tan \beta$ = 2, $\mu = 750$ GeV\@.  Dotted brown lines are contours of $\Omega_{\chi}^{(\rm th)} / \Omega_{\rm obs}$.   The gray regions are excluded by overabundance of neutralino dark matter, while the edge of this region has $\Omega_\chi^{(\rm th)} = \Omega_{\rm obs}$.  In the remainder of the plane $\chi$ is just one component of multi-component dark matter.  The present limit from XENON100 is shown shaded, while the projected reaches of LUX and XENON1T, both SI and SD, are shown as dashed lines.}
\end{figure}

Finally, non-thermal wino-like DM is also constrained by Fermi gamma ray searches.  The blue region of \Fig{fig:BinoWinoLimit} shows that dark matter being composed of winos is excluded up to a mass of about 500 GeV\@.  Compared to the limits described in \Sec{sec:bh}, indirect detection constraints are more stringent for wino-like DM than they were for Higgsinos due to group-theoretic factors.  As before, Fermi is currently unable to constrain thermal dark matter, which has a smaller annihilation cross-section.  For prior applications of Fermi limits to non-thermal wino DM, see for example~\cite{Ibe:2012hu, Belanger:2012ta, Hall:2012zp}.

The right-hand plot of \Fig{fig:BinoWinoLimit} is the same as the left except it shows future reach rather than current limits.  Here we see that XENON1T will exclude large swaths of dark matter, leaving small patches corresponding to SI blind spots (aside from a small region of very light non-thermal bino DM, which we will not consider further).  In particular, the horizontal and vertical black dashed lines denote the blind spots for bino-like and wino-like DM, respectively.  Consistent with \Eq{eq:SIblind}, for larger values of $\tan \beta$, these blind spots arise at smaller and smaller values of $|M_1|$ and $|M_2|$, respectively.  Spin-dependent direct detection limits, while not absent, are subdominant to SI limits throughout the plane of \Fig{fig:BinoWinoLimit} and are therefore not shown.

The contrast between the right and left plots is stark: currently TeV-scale neutralino dark matter is poorly constrained by direct detection experiments; but over the coming few years we can expect a much deeper probe of the theory yielding a large discovery region.   The absence of a signal will require either that parameters lie close to a blind spot or that mass parameters are above the TeV scale.   These conclusions persist with slices through the parameter space at other values of $\mu$ and $\tan \beta$. 

The diagonal black dashed line denotes the $M_1 = M_2$ blind spot for bino/wino DM\@.  Interestingly, this cancellation region coincides with the parameter space which accommodates a thermal relic abundance.  In the event that the bino/wino DM abundance is anthropically selected, the fine-tuning imposed to acquire the correct relic abundance automatically induces a cancellation in the SI scattering cross-section of dark matter---no additional tuning is required.

\subsection{Multi-Component Dark Matter with $\Omega_\chi = \Omega_\chi^{(\rm th)} \leq \Omega_{\rm obs}$}

\Fig{fig:BinoWinoSubdominant} shows the same slice through parameter space as Figure (\ref{fig:BinoWinoLimit}), so that the nature of the LSP in the various regions is as before.  The key difference is that we now assume that the relic LSP abundance is given purely by thermal freeze-out.  Hence, in addition to regions excluded by LEP searches for charginos and chargino LSPs, there is also a large region excluded by the overproduction of dark matter---essentially the entirety of the space containing a dominantly bino LSP\@.   One quadrant is almost entirely excluded by these considerations, although a narrow strip close to thermal dark matter is allowed.  

Contours of $\Omega_{\chi}^{(\rm th)} / \Omega_{\rm obs}$ are shown as dotted brown lines.    Regions shaded red have been excluded by XENON100, while future projected reaches are shown by dashed lines.  Both the limits and the reaches are highly asymmetrical in the four quadrants, which can be understood from the locations of the blind spots.    For example, XENON1T can limit the fraction of neutralino dark matter to less than 1\% for $M_{1,2}>0$, but does not reach 10\% if $M_2$ is negative.

 \begin{figure}[t!]
\begin{center} 
\includegraphics[scale=.55]{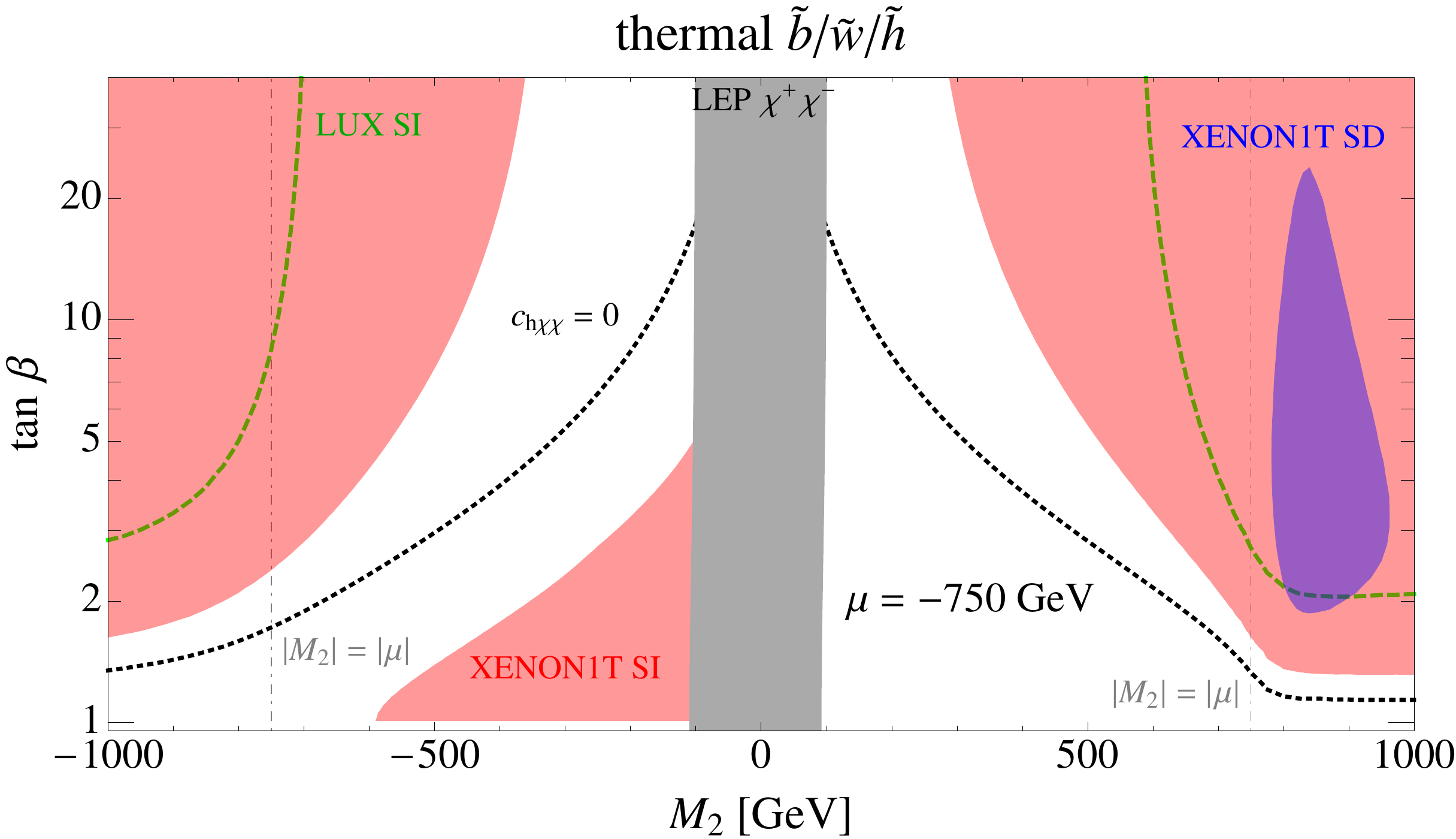}
\end{center}
\caption{\label{fig:BW_M2_v_tb_MuNeg}
Limit and reach for thermal bino/wino in the $M_2, \tan \beta$ plane for $\mu = -750$ GeV\@.  $M_1$ is fixed by requiring $\Omega_{\chi}^{\rm th} = \Omega_{\rm obs}$.  The black dotted lines correspond to the SI blind spot for bino-like DM given in \Eq{eq:binoblind}.  The XENON1T SI and SD exclusion reach is shown shaded in red and blue, respectively, while the LUX SI reach is shown with a dashed green line.  The LEP chargino exclusion is shaded in gray.}
\end{figure}
 
\subsection{Thermal Dark Matter with $\Omega_\chi = \Omega_\chi^{(\rm th)} =\Omega_{\rm obs}$}

We now restrict our attention to thermal bino/wino dark matter, which once again allows us to reduce the parameter space of interest by one dimension.   As before, we solve for $M_1$ using the thermal relic constant, assuming $M_1 > 0$.  We consider slices of the remaining parameter space at fixed $\mu$ and fixed $\tan\beta$.  \Fig{fig:BW_M2_v_tb_MuNeg} depicts limits and reach for thermal bino/wino dark matter in the $(M_2, \tan \beta)$ plane at $\mu = -750$ GeV\@.  The SI blind spot for bino-like dark matter is shown in the plot using the expression from \Eq{eq:binoblind}.

As discussed in the previous section, there is no current limit from XENON100 given this choice of the sign of $\mu$.  Furthermore, both LUX SI and XENON1T SD limits will only constrain the regions of parameter space with $|M_2| \gtrsim |\mu|$, which have a mixed bino/Higgsino LSP\@.  Spin-independent limits from XENON1T will cover much of the space with $M_2 < 0$, with the exception of a region around the SI blind spot.  For $M_2 > 0$, however, the reach is  weakened by virtue of the proximity of the $M_1 = M_2$ blind spot.  Note that no additional tuning is required beyond that which is needed to get the correct thermal relic abundance, only a discrete choice of sign.

\begin{figure}[t]
\begin{center} 
\hbox{ \hspace{-0.5cm}
\includegraphics[scale=.4]{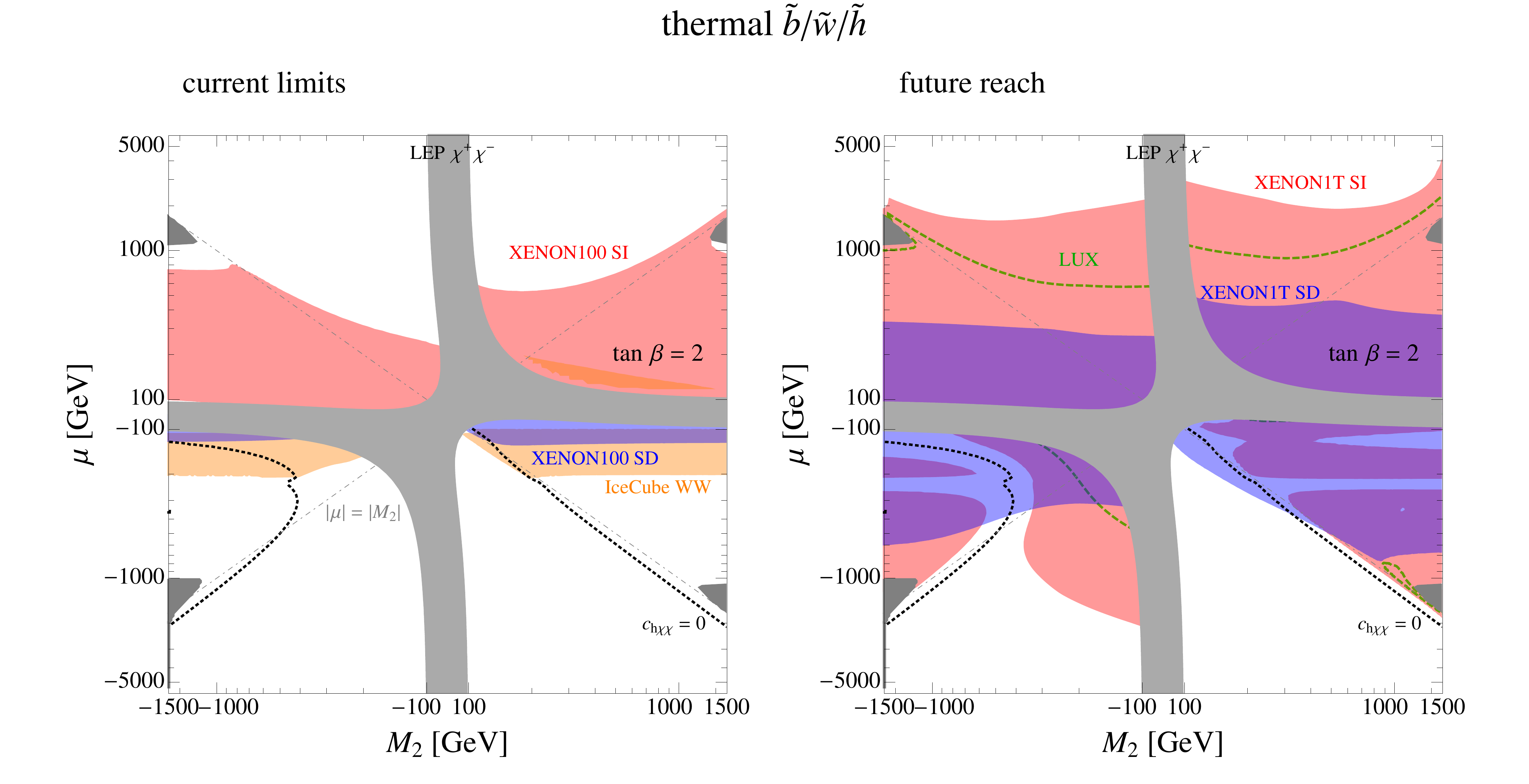} }
\end{center}
\caption{\label{fig:BW_M2_v_mu_limit}
Current limits and future reach for thermal bino/wino dark matter in the $(M_2, \mu)$ plane for $\tan\beta = 2$.  $M_1$ is fixed by requiring $\Omega_{\chi}^{\rm th} = \Omega_{\rm obs}$.  The black dotted lines correspond to the SI blind spot for bino-like DM given in \Eq{eq:binoblind}.  The XENON100 (XENON1T) SI and SD limits (reach) are shown shaded in red and blue, respectively on the left (right) side of the figure.  IceCube limits on dark matter annihilation to $W^+W^-$ are shown shaded in orange on the left, and the LUX SI exclusion reach is shown with a dashed green line in the right-hand panel.  The LEP chargino exclusion is shaded in light gray.  Darker gray regions correspond to overclosure via Higgsino-like DM heavier than 1 TeV.}
\end{figure}

Finally, \Fig{fig:BW_M2_v_mu_limit} shows the current limit and expected reach for thermal bino/wino dark matter in the $(M_2,\mu)$ plane at $\tan \beta = 2$.  Because of the location of the SI blind spots, \Fig{fig:BW_M2_v_mu_limit} depicts much weaker constraints for negative $\mu$ than positive $\mu$.  For small $\tan\beta$, the blind spots occur mostly in the bino/Higgsino region of the plane, in which $|M_2| \lesssim |\mu|$.  As $\tan\beta$ is raised, however the bind spots move to lower values of the gaugino mass relative to $\mu$, weakening the constraints on the bino/wino parameter space even further.   

Currently, there are no limits from SI direct detection for $\mu < 0$.  XENON1T will constrain bino/wino DM to lie near the $-M_1 = \mu \sin 2\beta$ blind spot for $\mu, M_2 < 0$; for $M_2 > 0$, however, the proximity of the well-tempered line to the $M_1 = M_2$ blind spot will once again weaken the constraints.  SD direct detection sets complementary limits, irrespective of the SI blind spots, but limits will remain relatively weak in the bino/wino region of the parameter space even after XENON1T because of the relatively small mixing angle.

 \begin{figure}[t]
\begin{center} 
\includegraphics[scale=.45]{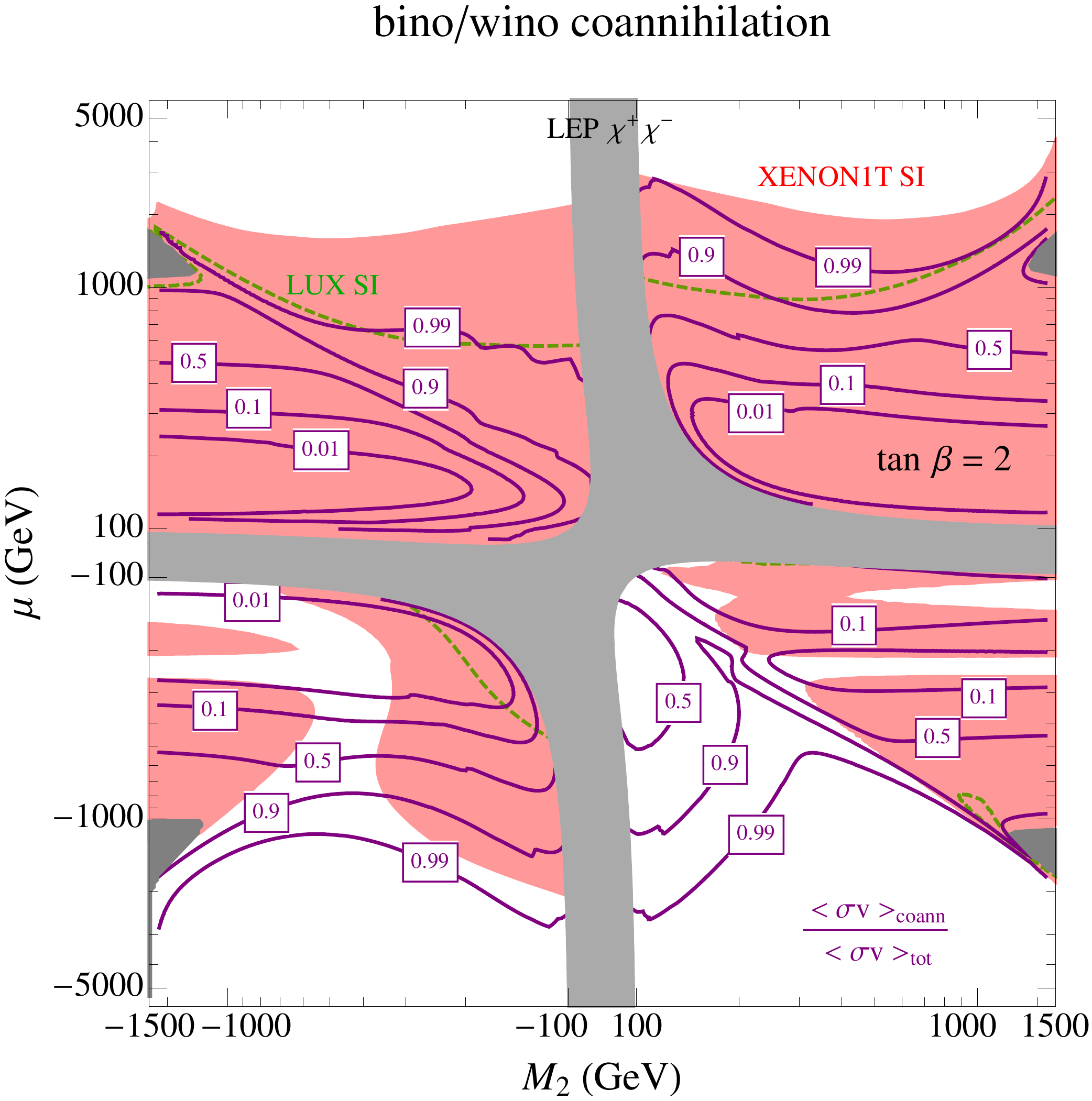}
\end{center}
\caption{\label{fig:BW_coA}
Contours of the fraction of dark matter annihilation cross-section, weighted by Boltzmann factors as \Eq{eq:coann}, coming from coannihilation are shown in purple, superimposed on the XENON1T SI exclusion reach shaded in red and the LUX SI reach in dashed green.  Once again, $\tan\beta$ is fixed to $2$, with $M_1$ fixed by requiring $\Omega_{\chi}^{\rm th} = \Omega_{\rm obs}$.}
\end{figure}

\begin{figure}[h]
\begin{center} 
\includegraphics[scale=.36 ]{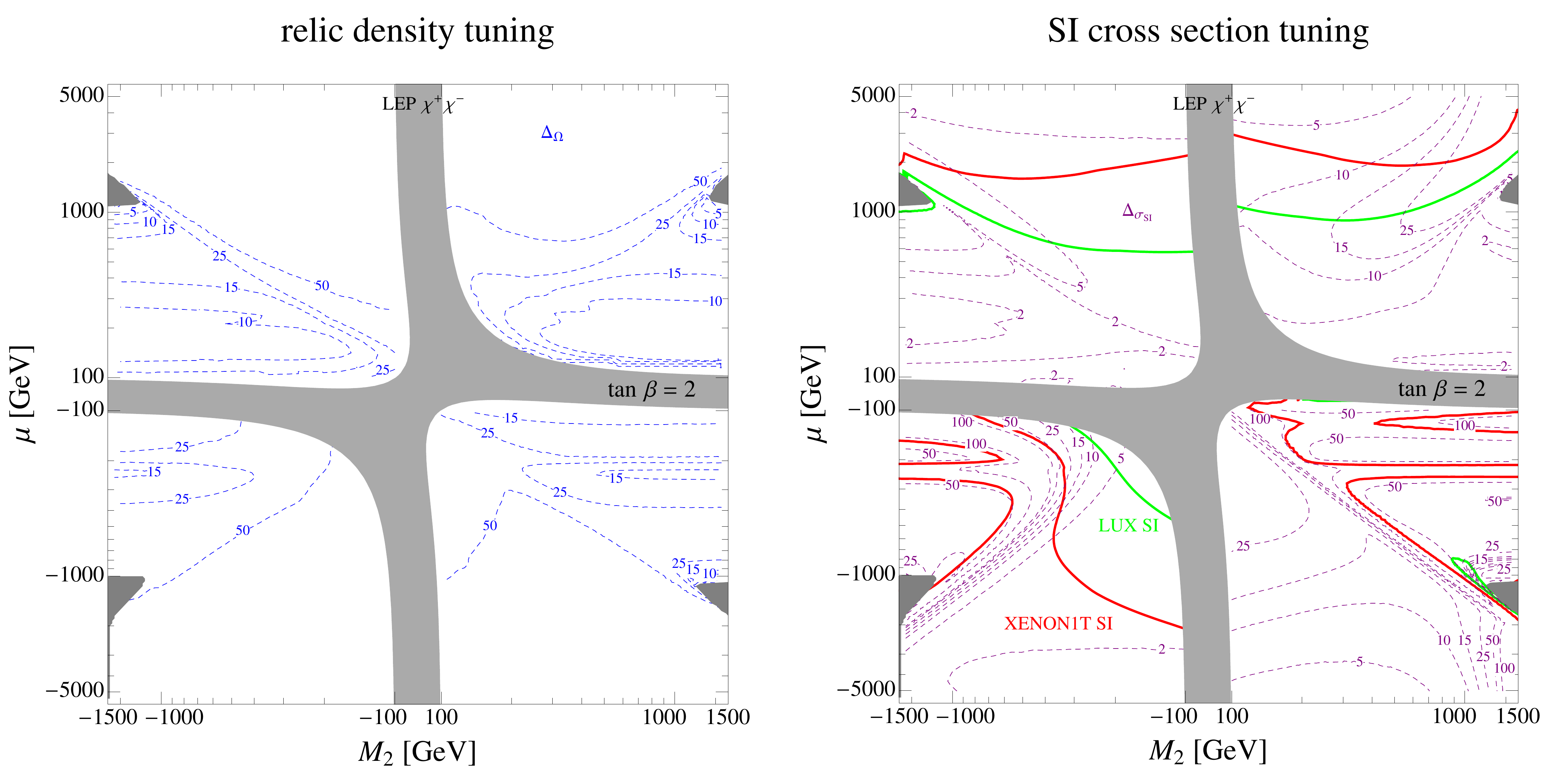}
\end{center}
\caption{\label{fig:BW_SI_tune}
Contours of fine-tuning of the relic abundance (SI scattering cross-section) are shown in the left (right) panel, using the measure defined in \App{app:tune}.  The XENON1T and LUX SI exclusion reaches are shown with solid red and green contours on the right.}
\end{figure}

Regardless of relative signs, the direct detection limits fall off as the magnitude of $\mu$ is raised, since both the Higgs and $Z$ couplings to dark matter are depleted as $\mu$ is decoupled.  Even at positive $\mu$, current limits from XENON100 only exclude $\mu \lesssim 250$ GeV, leaving large allowed regions with natural values of $\mu$ without requiring any tuning of the cross-section.  LUX and XENON1T will require $\mu \gtrsim 600$ and $1500$ GeV, respectively, when $\mu > 0$.  Evading these limits with large $\mu$ does not incur a fine-tuning penalty in the cross-section; however, as shown in \Fig{fig:BW_coA}, it is not without cost.  Bino/wino DM that evades LUX with $\mu > 0$ must rely heavily on coannihilation to achieve the observed relic density.  More than $90\%$ $(99\%)$ of the total dark matter annihilation cross-section, weighted by Boltzmann factors as in \Eq{eq:coann}, must come from coannihilation in order to escape LUX (XENON1T) limits with $\mu > 0$.    As discussed above, this coannihilation is exponentially sensitive to the mass difference between the bino and the wino, leading to a $\sim 2\%$ tuning in the relic abundance.  The left-hand panel of \Fig{fig:BW_SI_tune} shows how the relic abundance tuning increases as $\mu$ is decoupled.  The tuning increases as coannihilation becomes more important before reaching a plateau around $1/50$.

By way of contrast, the cross-section tuning, shown in the right-hand panel of \Fig{fig:BW_SI_tune}, takes relatively small values everywhere except close to the blind spots, where it quickly increases to $\sim 1\%$.  Note that in the lower right quadrant, it is possible to evade XENON1T with a relatively mild tuning of the cross-section, about $5\%$, and any value of $\mu$ consistent with the LEP bound, improving electroweak naturalness.

\section{Conclusions}
\label{sec:concl}

In this paper we have presented a systematic analysis of the current limits and projected reach for $\chi \sim ( \tilde b, \tilde w, \tilde h)$ neutralino DM using simplified models.  For bino/Higgsino and bino/wino(/Higgsino) DM we decoupled the heavier supersymmetric partners to yield a parameter space of $(M_1, \mu, \tan \beta)$ and $(M_1, M_2, \mu, \tan \beta)$, respectively.  We assumed CP conservation, but studied all physically distinct choices of signs of these parameters.  Furthermore, we assumed that the recently discovered state near 125 GeV is the Higgs boson, and that the other Higgs bosons are decoupled and yield subdominant contributions to the DM scattering cross-section.  We were then able to explore the current limits and future reach directly in this minimal parameter space without resorting to scatter plots.  This simplified model for analyzing neutralino DM is a good approximation to a wide variety of ultraviolet theories.  Limits and reaches were presented for neutralinos comprising all or just a fraction of dark matter.  The case of thermal freeze-out of neutralinos yielding all dark matter was emphasized, as this allows a reduction in the parameter space.
In what follows, we summarize out main results, and then discuss future directions.

 \begin{figure}[h!]
\begin{center} 
\includegraphics[scale=.55]{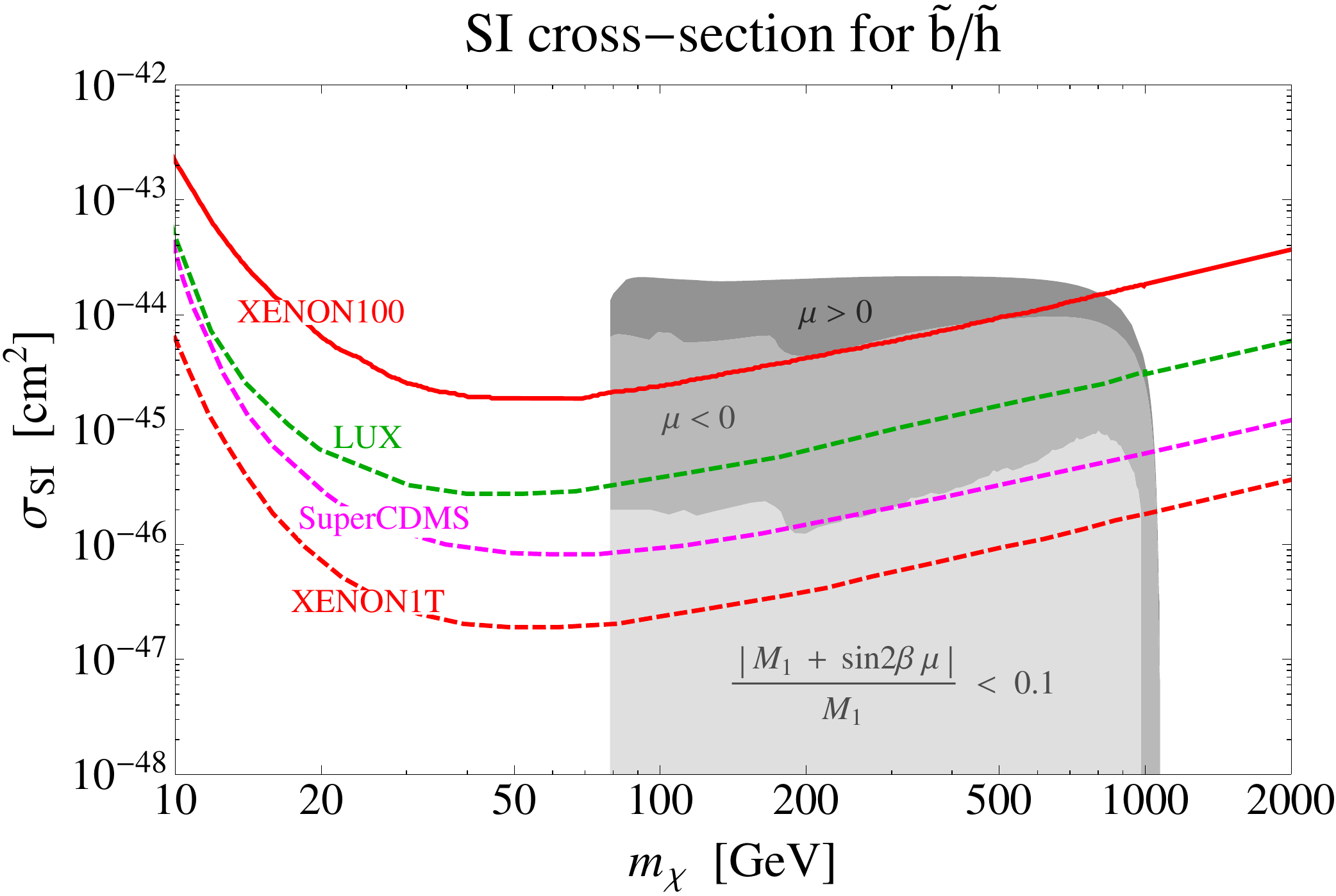}
\end{center}
\caption{\label{fig:amoeba}
The gray shaded areas depict target regions in the ($m_\chi, \sigma_{\rm SI}$) plane for thermal bino/Higgsino DM, superimposed on the current limit from XENON100 and the projected reaches for LUX and XENON1T\@.  The edge of these gray regions at low $m_\chi$ results from the LEP requirement of $|\mu| \gtrsim 100$ GeV, while the largest value of $m_\chi$, just above 1 TeV, corresponds to pure Higgsino LSP, and is present for both signs of $\mu$.  The upper dark shaded region is for $\mu >0$ (here we fix $M_1>0$) with the upper (lower) edge corresponding to low (high) $\tan \beta$. Much of the low mass part of this region has been excluded by XENON100.  The lower two regions, shaded in lighter gray, are for $\mu<0$.  The boundary between the $\mu>0$ and $\mu<0$ regions occurs at large $\tan \beta$, where the sign of $\mu$ becomes unphysical.  In the $\mu<0$ regions the cross-section falls as $\tan \beta$ is reduced towards its value at the blind spot, where $M_1 + \sin 2 \beta \, \mu = 0$.  The contour between the two $\mu <0$ regions is given by $|M_1 +  \mu  \sin 2 \beta| = 0.1\,  M_1$, roughly corresponding to a 10\% fine-tuning in the scattering amplitude.  In the lower region, for each order of magnitude further reduction in the cross-section, a factor of $\sqrt{10}$ more fine-tuning is required.  }
\end{figure}

Our results are detailed in \Fig{fig:BinoHiggsinoCross} - \Fig{fig:BW_SI_tune}, which depict present limits and future reach within the theory parameter space of neutralino DM\@.  However, direct detection experiments place bounds on the physical $(m_\chi,\sigma)$ plane.  For a proper comparison, \Fig{fig:amoeba}  depicts the image of thermal bino/Higgsino DM in the plane of physical parameters relevant to SI scattering.   The $\mu>0$ region has been excluded by XENON100 for values of $m_\chi$ up to about 500 GeV, but the $\mu<0$ region is almost entirely unconstrained.  The LUX, SuperCDMS and XENON1T experiments will probe this $\mu<0$ region deeply.  The absence of a signal would require a cancellation in the scattering amplitude at the level of 1 part in 10 - 30.

For non-thermal bino/Higgsino DM only a small fraction of parameter space with $|\mu|, M_1 < 1$~TeV has been excluded, as illustrated in \Fig{fig:BinoHiggsinoLimit}. A large (small) fraction of the parameter space for thermal DM has been excluded for $\mu>0$ ($\mu<0$), as can be seen most clearly in the upper panel of \Fig{fig:BH_WellTemper}.  Future experiments, such as LUX and XENON1T, have a large discovery potential, as they will explore the majority of the parameter space with $\mu, M_1$ up to 1 - 2 TeV, as shown in \Fig{fig:BinoHiggsinoReach}.  The hardest region to explore has low $\tan \beta$ and $\mu<0$, as this lies close to a blind spot, where $c_{h \chi \chi}=0$.  This is illustrated for thermal DM in the lower panels of \Fig{fig:BH_WellTemper}.  Pure Higgsino thermal dark matter will also evade discovery for $M_1 >2$ TeV, as shown by the vertical brown bands in \Fig{fig:BinoHiggsinoReach}.

\Fig{fig:multi} depicts current limits and projected reaches for bino/Higgsino LSP which is just one component of multi-component DM\@.  Present constraints are quite weak, but LUX and XENON1T will probe the fraction of LSP dark matter powerfully, especially at low LSP mass, although with the usual blind spot caveat at low $\tan \beta$.

The more general case of bino/wino/Higgsino DM is shown schematically in \Fig{fig:BinoWinoScheme}, and contains the interesting possibility of bino/wino thermal DM\@.  \Fig{fig:BinoWinoLimit} shows  the present limits and future reach for non-thermal production in a slice of parameter space.  While three of the four quadrants are affected by blind spots and are currently unconstrained by direct detection, all four quadrants will be significantly probed by XENON1T and LUX\@.   \Fig{fig:BinoWinoSubdominant} shows the same parameter slice for subdominant thermal DM where, depending on the quadrant, XENON1T can probe neutralinos that comprise 10\% or even 1\% of DM\@.  \Fig{fig:BW_M2_v_mu_limit} shows a slice at $\tan \beta = 2$ for the present limit and future reach for thermal DM\@.  For $\mu >0$ ($\mu <0$) the current limits are already quite strong (weak), but even the combination of both SI and SD XENON1T results will leave significant regions at $\mu<0$ with $|\mu|<$ 1 - 2 TeV\@.   In particular, the large open region in the lower right quadrant has $M_1$ close to $M_2$ from the requirement of coannihilation and hence lies close to the $M_1 = M_2$ blind spot, where $M_{1,2}$ have the opposite sign of $\mu$.

Finally, our results suggest a variety of interesting directions for future work, which we now summarize.  In particular,

\begin{itemize}

\item
The present analysis incorporates the critical effects of relative signs, but still assumes CP conservation in the neutralino sector.  The importance of these signs for alleviating direct detection limits suggest that similar effects can be expected of neutralino DM with arbitrary CP phases.  On the other hand, CP-violation in the neutralino sector is stringently constrained by electron electric dipole moment (EDM) experiments, subject to the masses of the scalar superpartners.  It would be interesting to study the interplay between DM direct detection and EDM experiments in CP-violating models of neutralino DM.  

\item
While the bulk of our analysis has been at tree level, our  estimates in \App{app:loop} suggest that they are robust to one loop corrections.  In particular, naive estimates indicate that the projected reach of XENON1T will cover parameter regions for which one loop corrections can be likely neglected.  However, an explicit calculation of one loop effects in the mixed bino/wino/Higgsino system will settle this issue, and have important implications for DM in the post-XENON1T era.

\item
In this work, we have primarily studied constraints from direct detection experiments and neutrino telescopes.  An interesting avenue for future work is a more comprehensive analysis of the indirect detection constraints on neutralino DM from cosmic ray and gamma ray observatories like the Fermi Telescope \cite{Ackermann:2011wa}.

\item
Our analysis applies to neutralino DM that is a general mixture of bino, wino, and Higgsino DM\@.  However, many of our qualitative results---{\it e.g}~the importance of relative signs---apply much more generally.  It would be very interesting to study present and future experimental results within the context of generic models of mixed singlet/triplet/doublet DM\@.   While simplified models of this type have been considered in a variety of contexts \cite{Cohen:2011ec, Mahbubani:2005pt, Essig:2007az,Elor:2009jp,Cheung:2012nb}, their interplay with present and future direct detection limits have been less systematically studied; in particular, the analogous blind spot parameter regions have not been fully identified.    A general singlet/triplet/doublet simplified model would provide a theoretically inclusive framework for studying---and, in the absence of positive signals, excluding---a huge class of WIMP DM models.  Such an analysis would have important implications for mixed singlino DM relevant to the NMSSM and its $\lambda$SUSY variants. 

\end{itemize}

\section*{Acknowledgments}

We thank Matthias Danninger, Beate Heinemann, Spencer Klein, Rafael Lang, Michele Papucci, and Satoshi Shirai for helpful conversations, and thank Gilly Elor for collaboration in the early stages of this work.  D.P.~is grateful to the Kavli IPMU for their hospitality during the completion of part of this work.  C.C, L.J.H, and J.T.R~would also like to thank the Aspen Center for Physics.  This work was supported in part by the Director, Office of Science, Office 
of High Energy and Nuclear Physics, of the US Department of Energy under 
Contract DE-AC02-05CH11231 and by the National Science Foundation under 
grants PHY-1002399 and PHY-0855653.  J.T.R. is supported by a fellowship from the Miller Institute for Basic Research in Science.

\appendix

\section{Strange Quark Content of the Nucleon}
\label{app:fs}

It is well known that direct detection limits are sensitive to nuclear physics uncertainties, in particular the strange squark content of the nucleon.  Higgs-mediated SI scattering is proportional to $|f|^2$, where \cite{Belanger:2008sj}
\be
f \equiv \frac{2}{9} + \frac{7}{9}\sum_{q = {u,d,s}} f_q \qquad \qquad f_q \equiv \frac{\langle N | m_q \bar{q} q | N \rangle}{m_N}.
\ee
The largest uncertainty comes from the strange quark content, $f_s$, since $f_u$ and $f_d$ are small, but different determinations have led to widely disparate values for $f_s$, as we now review.  Throughout, our analysis takes $f_u = f_d = 0.025$~\cite{Giedt:2009mr}.

 \begin{figure}[h!]
\begin{center} 
\includegraphics[scale=.575]{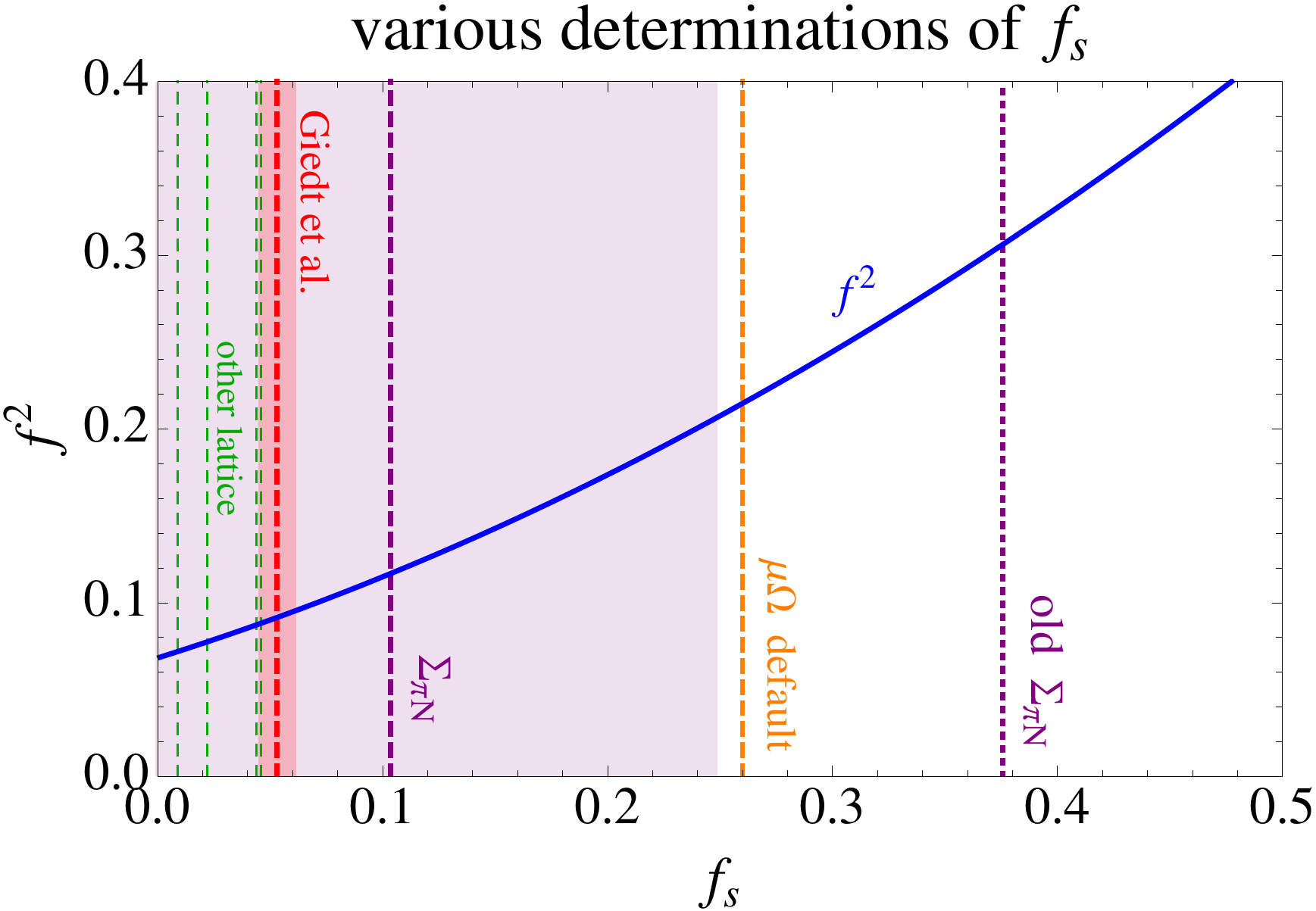}
\end{center}
\caption{\label{fig:fs}
The effect of $f_s$ on the Higgs-mediated SI cross-section, which is proportional to $f^2$.  We show the lattice determinations of $f_s$ from Giedt et al.~\cite{Giedt:2009mr} and more recent groups~\cite{Freeman:2012ry, Shanahan:2012wh,Oksuzian:2012rzb,Engelhardt:2012gd}, as well as the value of $f_s$ resulting from the most recent measurement of $\Sigma_{\pi N}$~\cite{Stahov}.  The bands indicate the stated $\pm 1 \sigma$ errors on the Giedt and $\Sigma_{\pi N}$ values, which are in agreement up to these errors.  For comparison, we show the default {\tt micrOMEGAs}~\cite{Belanger:2008sj} and traditional chiral perturbation theory~\cite{Ellis:2008hf} values, which are based on older measurements of $\Sigma_{\pi N}$.
}
\end{figure}

Traditionally~\cite{Ellis:2008hf}, $f_s$ was determined using chiral perturbation theory to relate $f_s$ to the pion-nucleon sigma term, $\Sigma_{\pi N}$, which is extracted experimentally from the cross-section for low-energy pion nucleon scattering, leading to $f_s = 0.38 \pm 0.10$.  The {\tt MicrOMEGAs}~\cite{Belanger:2008sj} default value, $f_s = 0.26$, is chosen to be near the $1\sigma$ lower bound based on these measurements of $\Sigma_{\pi N}$.  However, as first pointed out by Giedt et al.~\cite{Giedt:2009mr}, direct lattice determinations of $f_s$ lead to a significantly smaller value, $f_s = 0.0532 \pm 0.0085$.  More recent lattice results~\cite{Freeman:2012ry, Shanahan:2012wh,Oksuzian:2012rzb,Engelhardt:2012gd}  confirm small values, finding $f_s$ between $0.009 -0.046$.  This $\sim 3\sigma$ tension between the $\Sigma_{\pi N}$ and lattice determinations of $f_s$ have led to widely divergent approaches in the theory community, with some authors using large values of $f_s$ based on $\Sigma_{\pi N}$, and others adopting the lattice values.  We note that this tension is now probably resolved, in favor of the low values, because the most recent $p - \pi$ scattering data, from the CHAOS group at TRIUMF, lead to $f_s = 0.10 \pm 0.15$~\cite{Stahov}, in agreement with the lattice results.

In \Fig{fig:fs}, we show the effect on $f^2$ of the various choices for $f_s$ discussed above.  The cross-section does not change by a large amount between the different lattice determinations.  However, there is a large difference between the cross-section favored by the lattice and the cross-sections resulting from the old determinations of $\Sigma_{\pi N}$; for example the cross-section is increased by a factor of 2.3 when moving between the Giedt and default {\tt MicrOMEGAs} values. We view the lattice determination of $f_s$ to be most accurate (especially in light of the new, lower, value for $\Sigma_{\pi N}$) and throughout this paper we have used the value of Giedt et al.    We note that many previous theory studies (for example~\cite{Farina:2011bh,Perelstein:2011tg,Perelstein:2012qg}) have adopted the default {\tt MicrOMEGAs} value, leading to liberal limits by this factor of $\sim 2$.

\begin{figure}[h!]
\begin{center} 
\includegraphics[scale=.5]{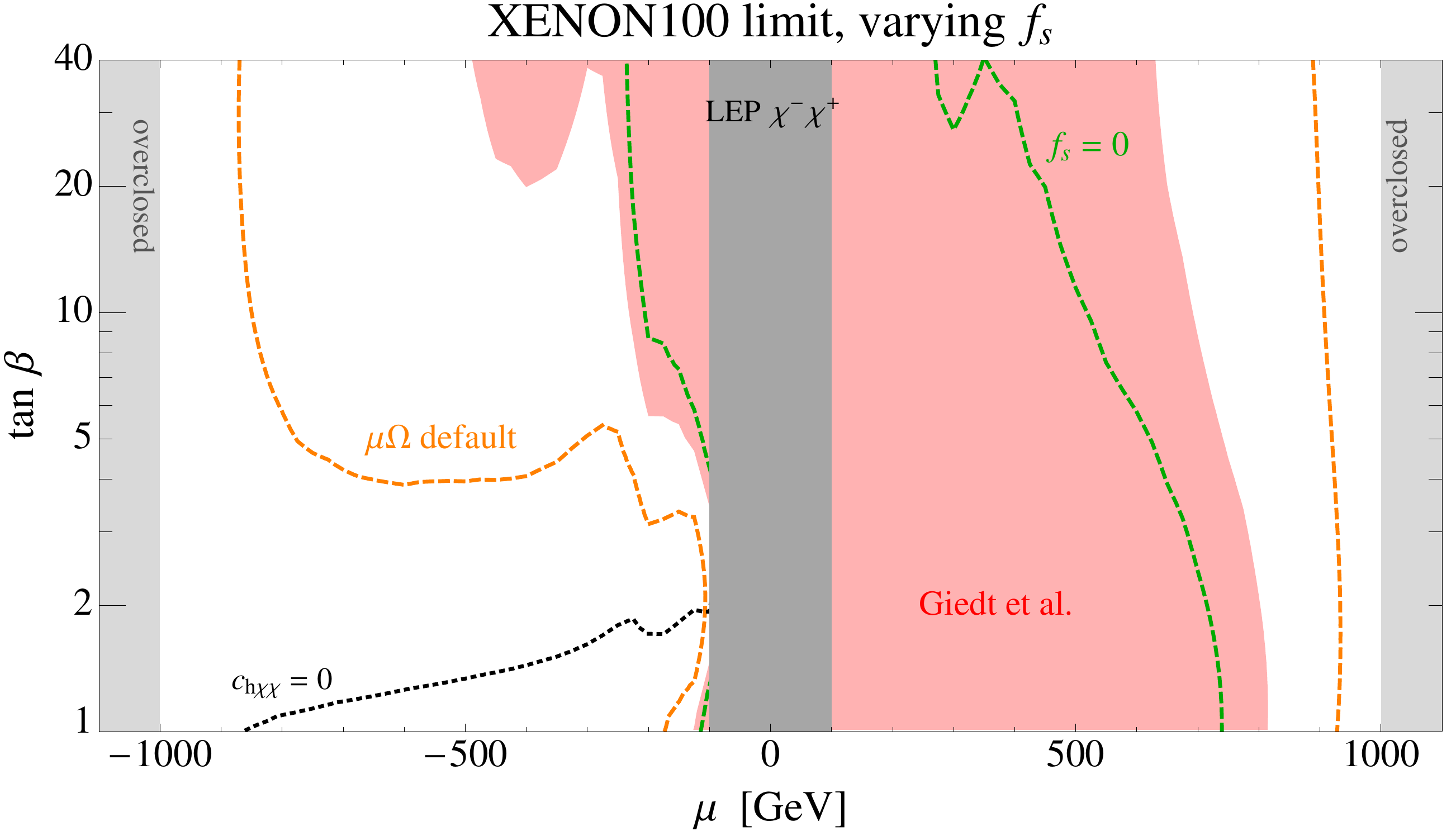}
\end{center}
\caption{\label{fig:Xenon100_fs}
The impact of different choices of $f_s$ on the XENON100 limit for thermal bino/Higgsino DM, which was shown using the Giedt et al. value in the first panel of \Fig{fig:BH_WellTemper}.  We compare the Giedt value to the default MicrOMEGAS value and $f_s = 0$, which is the most conservative choice.}
\end{figure}

We show the impact of varying $f_s$ on the XENON100 limit, for thermal bino/Higgsino DM, in \Fig{fig:Xenon100_fs}.  We compare the Giedt et al. value to the default {\tt MicrOMEGAs} value and to the most conservative possible choice of $f_s = 0$.  We see that the choice of $f_s$ makes a large difference at negative $\mu$ and large $\tan \beta$, because in this regime the cross-section is close to the XENON100 limit throughout much of the plane.  We comment that the thermal bino/Higgsino plane is the most sensitive of our results to the precise value of $f_s$, and the limit and reach contours throughout the rest of the paper are less sensitive to this choice.

\section{Dark Matter Fine-Tuning}
\label{app:tune}

In this paper we are agnostic about the possibility that the EW scale may be finely tuned.  But whether the weak scale is natural or not, it is interesting to identify when fine-tuning enters DM properties.  Such a tuning is a  worry when DM is well-tempered to produce the correct abundance, as discussed in \Sec{sec:relic}, or if DM sits particularly close to one of the blind spots we identified in \Sec{subsec:blindspots}.  In this appendix we describe a quantitative measure that tests the fine-tuning of $\Omega$ or $\sigma_{SI}$, independently of a possible EW tuning.  We applied the methodology described in this appendix to produce figures~\ref{fig:BH_Tuning} and \ref{fig:BW_SI_tune}, above.

To start, let us denote the log quantities, $p_i$ which label parameters at the weak scale: 
\bea
\exp p_i &\equiv & \{ M_1, M_2, \mu ,m_{H_u}^2, m_{H_d}^2, B\mu \}.
\eea
Then it is natural to define a log gradient defined as a directional directive with respect to log parameters, 
\bea
\vec \nabla \equiv  \nabla_i \equiv \frac{\partial}{\partial p_i}.
\eea
We can now define a vector in this space equal to the gradient of the electroweak symmetry breaking vacuum expectation value,
\bea
\vec V &\equiv & \vec \nabla \log v^2.
\eea
Here $\vec V$ is equal to the direction of steepest descent away from a particular value of $v$---thus, it is the combination of ultraviolet parameters which most strongly affects electroweak symmetry breaking.   In order to remove dependence on the possible fine-tuning of the electroweak symmetry breaking sector, we are interested in dependencies on parameters orthogonal to $\vec V$.  Explicitly, any $v$ dependent observable can be written as
\bea
\vec \nabla {\cal O}&=& \frac{\partial \cal O}{\partial \log v^2} \vec V +\ldots,
\eea
where we have used the chain rule and the ellipses denote dependencies on other parameters.  
Thus our fine-tuning parameter, which is independent of electroweak symmetry breaking, for a given observable ${\cal O}$ is defined by
\bea
\Delta_{\cal O }&\equiv & | \vec \nabla^{\perp} {\cal O}|,
\label{eq:finetuning}
\eea
where $\vec \nabla^{\perp}$ is defined as the gradient of ${\cal O}$ within the subspace of the above parameters orthogonal to $\vec V$.  
In  figures~\ref{fig:BH_Tuning} and \ref{fig:BW_SI_tune}, we used $\Delta_{\cal O}$, where ${\cal O} = \Omega, \sigma_{\rm SI}$, to quantify tuning in the thermal relic abundance,  and SI cross-section, respectively.  

We compute $\Delta_{\cal O}$ using  tree-level relationships among the weak scale parameters, with one exception.  The D-flat direction of the tree-level scalar potential causes the aforementioned tuning to blow up as $\tan\beta \rightarrow 1$ (see, for example, \cite{Perelstein:2011tg}).   We regard this apparently infinite tuning as unphysical, 
because the D-flat direction is lifted by loop corrections (and, moreover, whatever physics beyond the tree-level MSSM  accommodates the Higgs mass).
 In order to lift the D-flat direction, we assume an MSSM-like completion and add a quartic contribution $\propto |H_u|^4$, with a size fixed to reproduce $m_h = 125$~GeV\@.
The size of $\Delta_{\cal O}$ near $\tan\beta =  1$ depends on the exact form of potential;  however, any completion that lifts the D-flat direction serves to regularize the tuning as well.  We have verified that adding a quartic contribution of the form $|H_u H_d|^2$ yields similar results.

\section{Expected Size of Loop Corrections}
\label{app:loop}

To obtain our results we have only included the tree-level interactions of neutralino dark matter.  In most regions of parameter space this is a good approximation; however, close to the blind spots that we identified in \Sec{subsec:blindspots}, the relevant tree-level couplings to the Higgs and $Z$ bosons vanish, in which case it is crucial to ascertain the importance of one-loop corrections.  While a full computation of one-loop corrections is beyond the scope of the present work, the topic of this appendix is to estimate when loop corrections would modify our results for SI scattering.

 \begin{figure}[h!]
\begin{center} 
\includegraphics[scale=.475]{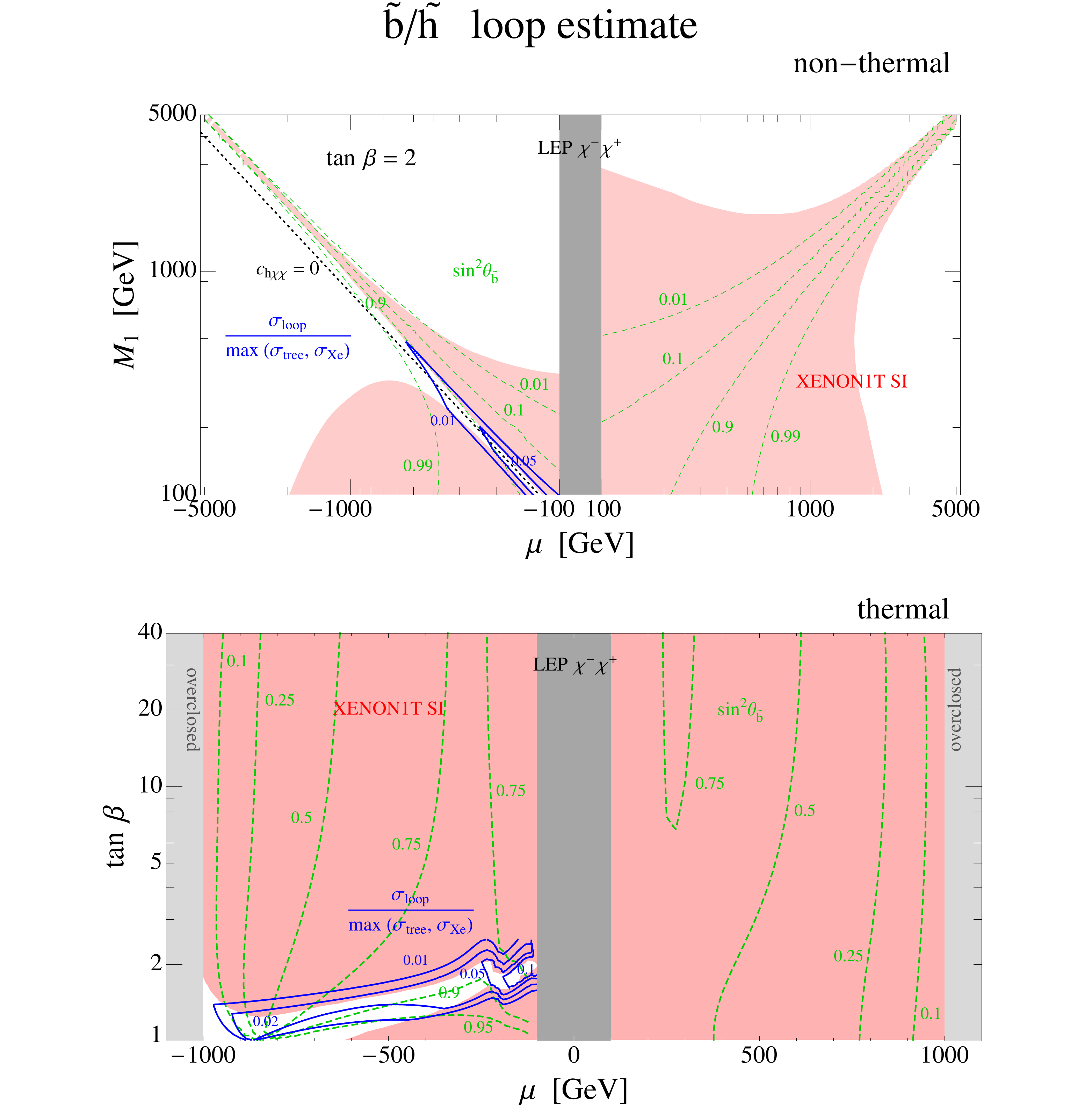}
\end{center}
\caption{\label{fig:BH_loop}
Estimated importance of loop corrections for bino/Higgsino DM, relative to the XENON1T reach.  Non-thermal and thermal DM are shown in the upper and lower panels, respectively, with the XENON1T reach as in \Fig{fig:BinoHiggsinoReach} (for $\tan \beta = 2$) and \Fig{fig:BH_WellTemper}.  The green contours show the bino fraction of DM, $\sin^2 \theta_{\tilde b} = Z_{11}^2$, and loop corrections are known to be small when DM is a pure bino or Higgsino, $\sin^2 \theta_{\tilde b} \approx 0, 1$. Blue contours show the ratio of our estimated loop cross-section to the sum of the tree-level and XENON1T reach cross-sections.   A large value of this ratio would indicate where the XENON1T reach estimate is sensitive to loop corrections, however we see that this ratio is small throughout most of parameter space, indicating that our results are robust to loop corrections.
}
\end{figure}

First, consider the case that DM is a pure eigenstate.  If DM is pure bino, it is inert (with decoupled scalars) and there are no radiative corrections to the Higgs coupling to the neutralino.  On the other hand, if DM is a pure Higgsino or pure wino, it has no Higgs coupling at tree-level, but there are loop diagrams such as box diagrams with two gauge bosons.  The naive size of these loop corrections is $10^{-(46-47)}~\mathrm{cm}^2$; large enough to probe at XENON1T\@.  However, a diagrammatic calculation including all one-loop and leading two-loop diagrams shows a surprising accidental cancellation among scalar and tensor scattering operators generated by one-loop and two-loop effects~\cite{Hisano:2011cs}.  For example, taking a pure Higgsino (wino) with a mass of 500 GeV, and using the form factors of~\cite{Giedt:2009mr}, the total cross-section is $6\times10^{-50} ~ (8 \times 10^{-48})~\mathrm{cm}^2$; much too small to probe at XENON1T\@.  Recently, this accidental cancellation was confirmed in an effective field theory calculation~\cite{Hill:2011be}.

The multiloop result of Refs.~\cite{Hisano:2011cs, Hill:2011be} have not yet been generalized to mixed states, as would be relevant for the blind spots.   We can still estimate the maximum size of the loop corrections.  Consider the case of mixed bino/Higgsino.  The bino component of the DM is given by $\sin^2 \theta_{\tilde b} = Z_{11}^2$, where $Z_{ij}$ is the rotation matrix going from interaction to mass eigenstate.  If $\sin^2 \theta_{\tilde b} \approx 0, 1$, then the DM is close to a pure state and the cross-section is small, as discussed above.  This quantity is shown by green curves for non-thermal and thermal cosmologies in the upper and lower panels of \Fig{fig:BH_loop}, respectively, and we see that much of the parameter space is characterized by a nearly pure state.  As the mixing angle is increased, any possible enhancement to the cross-section is suppressed at least by a factor of mixing angle squared.  Therefore, we conservatively estimate the maximum size of the cross-section to be, 
\bea
\sigma_{\rm loop} &=& |Z_{11}|^2 (1-|Z_{11}|^2) \times (2\times10^{-47}~\mathrm{cm}^2).
\label{eq:loopest}
\eea
Here, the number in parentheses reflects the size of the cross-section from the largest individual loop diagram contributing to the Higgsino cross-section from Ref.~\cite{Hisano:2011cs} (which happens to be a box of $W$ bosons contributing to the tensor operator).  We stress that this is simply an estimate of the maximum size; the full calculation is beyond our scope.

When are our results sensitive to the loop corrections?  Consider the XENON1T reach, shown for bino/Higgsino in \Fig{fig:BH_loop}.  For the loop correction to be relevant, two conditions must be satisfied, (1) the loop contribution must be large relative to the tree-level scattering, and (2) the loop contribution must be large enough to probe at XENON1T\@.  In order to estimate when both of these conditions are met, the blue contours in \Fig{fig:BH_loop} show the ratio of $\sigma_{\rm loop}$ to the maximum of the tree-level cross-section and the XENON1T limit.  The loop correction is important when this quantity is large.   However, this quantity is less than 0.01 in the entire parameter space, except for small regions near the blind spot.  Even here, it is only at low $\mu$ that this ratio reaches 0.05, and its maximum value is near 0.2.  Hence, unless our estimate \Eq{eq:loopest} is too small by over an order of magnitude, the loop corrections can be ignored for determining the reach of 1T detectors for bino/Higgsino DM.

We have not included an estimate of the importance of loop corrections for mixed DM with a large wino component.  In this case there are competing effects; the bino/wino DM typically has a smaller mixing angle than the bino/Higgsino case, further suppressing any enhancement to the loop contribution coming from mixing, but the largest individual loop diagram for pure wino scattering is an order of magnitude larger than for the pure Higgsino case.


\end{document}